\documentclass[pra,aps,superscriptaddress,twocolumn,nofootinbib, accepted=2019-09-16]{quantumarticle}
\pdfoutput=1
\usepackage[utf8]{inputenc}
\usepackage[english]{babel}
\usepackage[T1]{fontenc}
\usepackage{amsmath}
\usepackage{hyperref}

\usepackage{tikz}
\usepackage{lipsum}
\usepackage{amsmath}
\usepackage{amssymb}
\usepackage{epsfig}
\usepackage{color,epstopdf}
\usepackage{amsthm, thm-restate}
\usepackage{hyperref}
\usepackage{soul}
\usepackage[caption=false]{subfig}
\usepackage{algpseudocode}

\newtheorem*{theorem*}{Theorem}

\newtheorem{lemma}{Lemma}
\newtheorem*{lemma*}{Lemma}

\newtheorem{definition}{Definition}

\newcommand{\ket}[1]{|#1\rangle}
\newcommand{\bra}[1]{\langle #1|}

\newcommand{\ketbra}[2]{|#1\rangle\langle #2|}
\newcommand{\tr}[0]{\textnormal{Tr}}

\renewcommand{\v}[1]{\ensuremath{\boldsymbol #1}}

\begin{document}
	
	\title{Heat-Bath Algorithmic Cooling with optimal thermalization strategies}
	\author{\'Alvaro M. Alhambra}
	\affiliation{Perimeter Institute for Theoretical Physics, Waterloo, ON N2L 2Y5, Canada}
	\author{Matteo Lostaglio}
	\affiliation{ICFO-Institut de Ciencies Fotoniques, The Barcelona Institute of Science and Technology, Castelldefels (Barcelona), 08860, Spain}  
	\author{Christopher Perry}
	\affiliation{QMATH, Department of Mathematical Sciences, University of Copenhagen, Universitetsparken 5, 2100 Copenhagen, Denmark}
	
	\maketitle
	
	\begin{abstract}
		Heat-Bath Algorithmic Cooling is a set of techniques for producing highly pure quantum systems by utilizing a surrounding heat-bath and unitary interactions. These techniques originally used the thermal environment only to fully thermalize ancillas at the environment temperature. Here we extend HBAC protocols by optimizing over the thermalization strategy. We find, for any $d$-dimensional system in an arbitrary initial state, provably optimal cooling protocols with surprisingly simple structure and exponential convergence to the ground state. Compared to the standard ones, these schemes can use fewer or no ancillas and exploit memory effects to enhance cooling. We verify that the optimal protocols are robusts to various deviations from the ideal scenario. For a single target qubit, the optimal protocol can be well approximated with a Jaynes-Cummings interaction between the system and a single thermal bosonic mode for a wide range of environmental temperatures. This admits an experimental implementation close to the setup of a micromaser, with a performance competitive with leading proposals in the literature. The proposed protocol provides an experimental setup that illustrates how non-Markovianity can be harnessed to improve cooling. On the technical side  we 1. introduce a new class of states called \emph{maximally active states} and discuss their thermodynamic significance in terms of optimal unitary control, 2. introduce a new set of thermodynamic processes, called \emph{$\beta$-permutations}, whose access is sufficient to simulate a generic thermalization process, 3. show how to use abstract toolbox developed within the resource theory approach to thermodynamics to perform challenging optimizations, while combining it with open quantum system dynamics tools to approximate optimal solutions within physically realistic setups. 
	\end{abstract}
	
 Cooling is a central problem in quantum physics and in realizing technologies for quantum information processing. The ability to produce a set of highly pure, `cold', quantum states is vital for the construction of a quantum computer \cite{divincenzo2000physical}. More generally, the observation of quantum effects often requires cooling and, as such, many techniques have been developed to cool systems efficiently in platforms ranging from cavity optomechanics \cite{liu2013dynamic} to NMR \cite{schulman1999molecular, boykin2002algorithmic}, ion traps \cite{eschner2003laser} and superconducting qubits~\cite{valenzuela2006microwave}.

Here we use powerful techniques, developed within the resource theory approach to thermodynamics \cite{lostaglio2018thermodynamic}, to greatly extend an important class of cooling algorithms known as Heat-Bath Algorithmic Cooling (HBAC) \cite{schulman1999molecular,boykin2002algorithmic}. The goal of these is to maximize the purity of a target system $S$ in a given number of cooling rounds. Each round of the algorithm starts with a unitary applied to the target together with several auxiliary systems $A$ initialized in a thermal state, with the aim of pumping entropy away from the target. Next, the auxiliary systems are re-thermalized through coupling with a heat-bath, before the entire process is repeated in the next round. The asymptotically optimal protocol of this form (in terms of the purity reached in infinitely many rounds) is the Partner Pairing Algorithm (PPA), introduced in~\cite{schulman2005physical}, whose asymptotic performance has been recently derived for a single target qubit starting in a maximally mixed state~\cite{rodriguez2016achievable,raeisi2015asymptotic}. 

Here, we generalize HBAC in the following sense: instead of using the thermal environment only to refresh the auxiliary systems through a \emph{complete} thermalization, we allow strategies involving an \emph{incomplete} thermalization of system and ancillas. In particular, we optimize the protocol over \emph{every} `dephasing thermalization', that is any quantum map on $SA$ which 1. leaves the thermal state on $SA$ fixed and 2. dephases the input state in the energy eigenbasis. This generalizes the `rethermalization' used in previous protocols, and thus extends HBAC to a larger set that will be called Extended Heat-Bath Algorithmic Cooling (xHBAC).  For example, the recent protocol introduced in Ref.~\cite{rodriguez2017heat}, which proposes the use of the heat bath to implement a non-local thermalization  `state-reset' (SR) process related to the Nuclear Overhauser Effect \cite{overhauser1953aw}, can already be seen as a (non-optimal) protocol within our extended family of xHBAC.

For \emph{every} finite dimensional target state in an \emph{arbitrary} initial state, we optimize the cooling performance over \emph{every} xHBAC for \emph{any} given number of rounds. We give an analytical form for the optimal cooling operations, uncovering their elegant structure, and show that the ground state population goes to $1$ exponentially fast in the number of rounds. This opens up a new avenue in the experimental realization of algorithmic cooling schemes, one requiring control over fewer ancillas, but better control of the interaction with the environment.

Our results suggest that xHBAC schemes provide new cooling protocols in practically relevant settings. We show that, for a single qubit target and no ancillas, the optimal protocol in xHBAC can be approximated by coupling the qubit by a Jaynes-Cummings (JC) interaction to a single bosonic thermal mode, itself weakly coupled to the external bath. The memory effects present in the JC interaction are crucial in approximating the theoretical optimal cooling. The performance of the ideal cooling protocol is robust to certain kinds of noise and imperfections, and it outperforms HBAC schemes with a small number of auxiliary qubits. This makes it, in our opinion, the most promising proposal for an experimental demonstration of cooling through xHBAC, as well as an experimental demonstration that non-Markovianity can be harnessed to improve cooling.

\section{Results and discussion}

\subsection{A general cooling theorem}\label{sec:general}

\begin{figure}[t!]
	\includegraphics[width=1\linewidth]{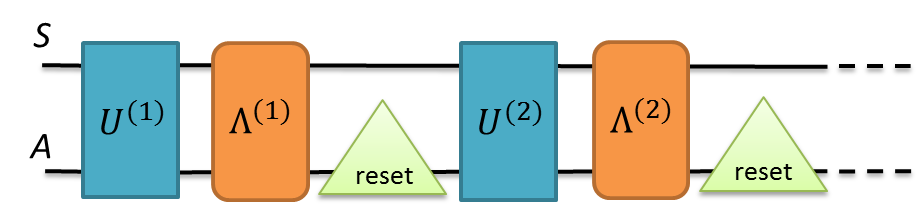}
	\caption{\small A generic xHBAC protocol.}
	\label{fig:generalprotocol}
\end{figure}

A general xHBAC will consist of a number of rounds and manipulate two types of systems, the target system $S$ to be purified and the auxiliary systems $A$. Furthermore, we will assume we can access a thermal environment at inverse temperature $\beta = (kT)^{-1}$, with $k$ Boltzmann's constant.
We denote by $H_{S}=\sum_{i=0}^{d-1} E_i \ketbra{i}{i}$, $E_0 \leq \dots \leq E_{d-1}$, the Hamiltonian of $S$ and by $\rho^{\left(k\right)}_S$ the state after round $k$. Also, we denote by $H_A$ the Hamiltonian of the auxiliary systems, initially in state $\rho_A$, and by $\tau_X = e^{-\beta H_X}/\tr{[e^{-\beta H_X}]}$ the thermal state on $X=S,A$.
A protocol is made of $k$ rounds, each allowing for (Fig.~\ref{fig:generalprotocol}):
\begin{enumerate}
	\item \textbf{Unitary.} Any unitary $U^{(k)}$ applied to $SA$.
	
	\item \textbf{Dephasing thermalization.} Any $\Lambda^{(k)}$ applied to $SA$, where $\Lambda^{(k)}$ is any quantum map such that \emph{i)} $\Lambda^{(k)}(\tau_{S} \otimes \tau_{A})=\tau_{S} \otimes \tau_{A}$ (thermal fixed point); \emph{ii)} If $\ket{E}$, $\ket{E'}$ are states of distinct energy on $SA$, $\bra{E} \Lambda^{(k)}(\rho_{SA}) \ket{E'} = 0$ (dephasing).
\end{enumerate}

At the end of the round, a refreshing stage returns the auxiliary systems $A$ to their original state $\rho_A$. We denote the set of protocols whose rounds have this general form by $\mathcal{P}_{\rho_A}$. Typically $\rho_A = \tau_A$, and this operation is simply a specific kind of dephasing thermalization. However, more generally, $\rho_A$ may also be the output of some previous cooling algorithm, subsequently used as an auxiliary system. Note that $A$ is distinguished by the rest of the environment in that it is under complete unitary control.

Some comments about xHBAC protocols are in order. First, if we skipped the dephasing thermalizations and $\rho_A = \tau_A$, we would get back to a standard HBAC scheme. 
xHBAC protocols include thermalization of subsets of energy levels, as in the mentioned SR protocol of Ref.~\cite{rodriguez2017heat}. But they are by no means limited to these strategies. Second, for every $k$ we
wish to find a protocol maximizing the ground state population $p^{\left(k\right)}_0=\bra{0}\rho^{\left(k\right)}_S\ket{0}$ over all sequences of $k$-round operations. Note that often the literature on cooling is restricted to finding asymptotically optimal protocols (e.g., the PPA~\cite{schulman2005physical}), but here we find optimal $k$ rounds protocols for every $k$. Finally, we note in passing that within this work we will not make use of auxiliary `scratch qubits'~\cite{rodriguez2016achievable,raeisi2015asymptotic}, i.e. $S$ itself is the target system to be cooled.

To give the analytical form of the optimal protocols we need to introduce two notions. First, that of \emph{maximally active} states. Given a state $\rho$ with Hamiltonian $H$, the maximally active state $\hat{\rho}$ is formed by diagonalizing $\rho$ in the energy eigenbasis and ordering the eigenvalues in increasing order with respect to energy. That is, $\hat{\rho}$  is the most energetic state in the unitary orbit of $\rho$. 

Second, we define the \emph{thermal polytope} in the following way. Given an initial state $\rho$ with populations in the energy eigenbasis $\textbf{p}$, it is the set of 
populations $\v{p}'$ of $\Lambda(\rho)$, with $\Lambda$ an arbitrary dephasing thermalization. Without loss of generality, we assume that every energy eigenspace has been diagonalized by energy preserving unitaries.
The thermal polytope is convex and has a finite number of extremal points (Lemma~12, \cite{lostaglio2018elementary}). It is particularly useful in thermodynamics to classify out of equilibrium distributions according to their $\beta$-order~\cite{horodecki2013fundamental}. Given a vector $\v{p}$ and a correspondent Hamiltonian with energy levels $E_i$, the $\beta-$order of $\v{p}$ is the permutation of $\pi$ of $0, 1, 2, \dots$ such that the Gibbs-rescaled population is sorted in non-increasing order:  $p_{\pi(0)} e^{\beta E_{\pi(0)}} \geq p_{\pi(1)} e^{\beta E_{\pi(1)}} \geq \dots$.

\begin{figure}[t!]
	\includegraphics[width=1\linewidth]{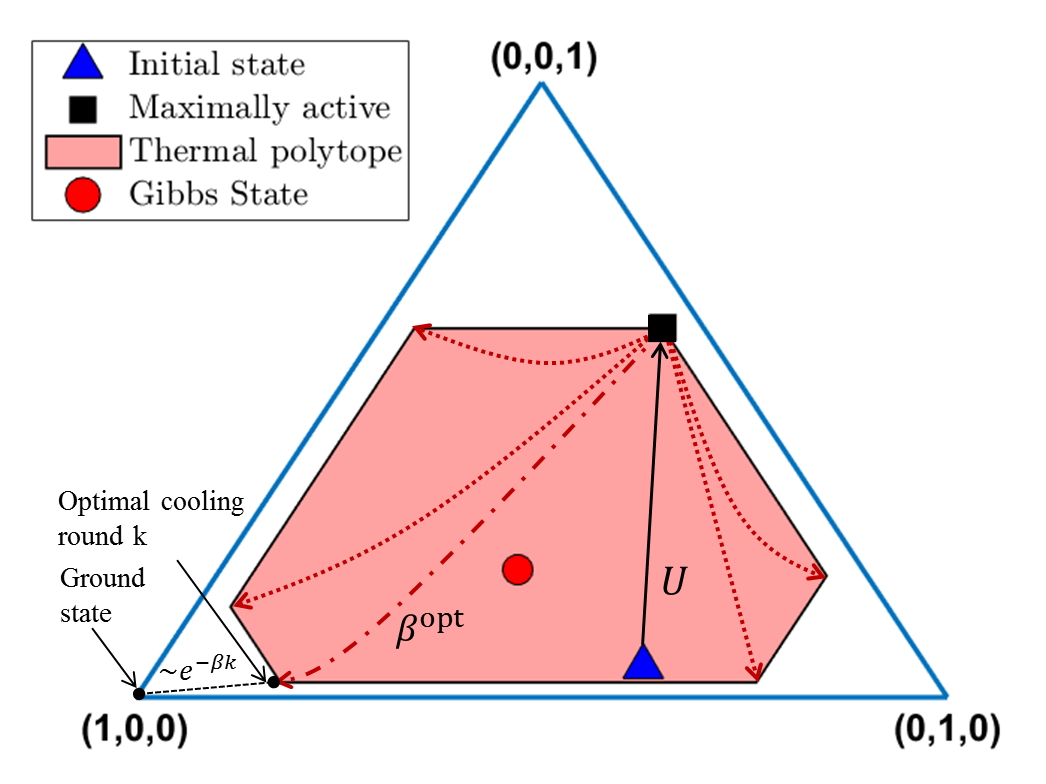}
	\caption{\small An optimal cooling round on a $d=3$ system. Map the initial state after round $k-1$ to its maximally active state by a unitary $U$ (continuous black arrow). The thermal polytope of the maximally active state then describes all possible final states achievable by arbitrary thermalizations. $\beta$-permutations map to the extremal points of these polytope (dotted red arrows), and $\beta^{\rm opt}$ achieves the one closest to the ground state (dash-dot red arrow). The distance to the ground state decreases exponentially fast in the number $k$ of rounds.}
	\label{fig:polytope}
\end{figure}

Crucially, for any dimension $d$, we provide an explicit construction of a set of dephasing thermalizations, denoted $\Lambda^{(\pi,\alpha)}$ and called $\beta$-\emph{permutations}, able to map any given $\v{p}$ to every extremal point of its thermal polytope (Methods, Sec.~\ref{ap:thermalpolytope} and Fig.~\ref{fig:polytope}). 
$\pi$ and $\alpha$ are two vectors of integers, each ranging over all permutations of $\{0,...,(d-1)(r-1)\}$, if $r$ is the dimension of $A$. Their significance is the following. If we are given an initial state $\v{p}$ with $\beta$-order $\pi$, the state $\v{q}^\alpha := \Lambda^{(\pi, \alpha)}(\v{p})$ is the `optimal' state among all those with $\beta$-order $\alpha$. More precisely, there is no other state in the thermal polytope of $\v{p}$ which has $\beta$-order $\alpha$ and can be transformed into $\v{q}^\alpha$ by dephasing thermalizations. This is in analogy with the role of permutations under doubly stochastic maps. In fact, in the limit $\beta \rightarrow +\infty$, $\beta$-permutations are permutations and the $\beta$-ordering is the standard sorting.

A particularly important $\beta$-permutation, denoted by $\beta^{\rm opt}$, is the one that maximizes the ground state population of $S$ among all dephasing thermalizations and, furthermore, achieves the largest partial sums $\sum_{i=0}^l p^{(k)}_i$, $l=0,...,d-1$, $p^{(k)}_i = \bra{i}\rho^{(k)}_S \ket{i}$. $\beta^{\rm opt}$ is constructed as follows: $ \beta^{\rm opt}= \Lambda^{(\pi,\alpha)}$ with $\pi$ an ordering of energy levels of $SA$ such that, if $q^{(k)}_m$ are populations of $SA$ at round $k$ and $E^{SA}_m$ are the energies of $H_S + H_A$, $q^{(k)}_{\pi(m)} e^{\beta E^{SA}_{\pi(m)}}$ are sorted in non-increasing order in $m$; and 
\begin{align}
\alpha= &\big\{\left(0,r-1\right), \nonumber
\left(0,r-2\right),\dots,\left(0,0\right), \\& \left(1,r-1\right),\left(1,r-2\right),\dots,\left(1,0\right),\dots,\left(d-1,0\right)\big\}. \label{eq:alpha}
\end{align} 

This identifies an optimal dephasing thermalization. But what is the optimal unitary control that we need to apply? The following lemma answers this question:
\begin{lemma}\label{le:active}
	Given a state $\rho$ with population $\v{p}$, the thermal polytope of the maximally active state $\hat{\rho}$ contains the thermal polytope of $U \rho U^\dag$ for every unitary $U$.  
\end{lemma}

The proof can be found in the Methods, Sec.~\ref{ap:lemmaoptimalcontrol}. With these concepts in place, we can give the optimal cooling protocol for any given set of auxiliary systems in $A$.
\begin{restatable}{theorem}{thdlevel} \label{th:dlevel}
	Assume $S$ is a $d$-level system with \mbox{$E_{d-1}>E_0$} and $\beta>0$. Without loss generality (by an initial diagonalizing unitary) we can take the initial state of $S$ to be diagonal in the energy eigenbasis with $p^{\left(0\right)}_0\geq\dots\geq p^{\left(0\right)}_{d-1}$. Then, for a given auxiliary state $\rho_A$, the optimal cooling protocol in $\mathcal{P}_{\rho_A}$ is such that in each round~$k$:
	\begin{enumerate}
		\item The unitary mapping $\rho^{\left(k-1\right)}_{S}\otimes\rho_A$ to the corresponding maximally active state is applied to $SA$.
		\item $\beta^{\rm opt}$ is applied to $SA$.
	\end{enumerate}
	The optimal protocol achieves $p_0^{(k)} \rightarrow 1$ at least exponentially fast in $k$, even with no ancilla.
\end{restatable} 

The intuition behind this protocol is simple (see  Fig.~\ref{fig:polytope}). In a single round, the unitary maximizes the amount of energy in $SA$, in accordance to Lemma~\ref{le:active}. Next, the optimal $\beta$-permutation is applied. For the proof, see Methods (Sec.~\ref{ap:prooftheorem1}).

In analyzing the performance of the protocols $\mathcal{P}_{\rho_A}$ it is important to keep in mind the cost of preparing and controlling the auxiliary systems $A$, especially if $\rho_A \neq \tau_A$. Remarkably, however, the optimal xHBAC protocol that uses no auxiliary systems $A$ still has $p^{(k)}_0 \rightarrow 1$ exponentially in $k$. Furthermore, such protocol has the further advantage that the cooling operations do not change with the round $k$. Also note that the initial state-dependent unitary can always be replaced by a complete thermalization while maintaining the same cooling scaling, so that no prior knowledge of $\rho^{(0)}_S$ is needed to unlock a strong cooling performance. Finally, in the Methods section~\ref{ap:protocol}, we construct for any $d$ an explicit protocol that not only uses no ancillas, and can be realized by a sequence of two level interactions: first, a unitary swaps the populations of the ground and most excited states; then, a sequence of two-level dephasing thermalizations between energies  $(d-1,d-2)$, $(d-2,d-3)$, $\dots$, $(1,0)$ is performed, each maximising the net population transfer $i \mapsto i-1$ (these are $\beta$-swaps, discussed in more detail in the next section). Every  $d-1$ repetitions, the performance takes an elegant form:
\begin{equation}
\label{eq:ddimensional}
p^{\left(k \left(d-1\right)\right)}_0 = 1- e^{-k \beta \left(E_{d-1}- E_0\right)}\left(1- p^{\left(0\right)}_0\right).
\end{equation}

\subsection{The qubit case}

To show how the general results can be used in a specific case, let us analyze in detail the single qubit case with no auxiliary systems $A$ (denoted $\mathcal{P}_\emptyset$) and energy gap $E$. The only nontrivial $\beta$-permutation is $\Lambda_\beta = \Lambda^{(\pi,\alpha)}$ with \mbox{$\pi = \{1,0\}$} and \mbox{$\alpha = \{0,1\}$}, which induces transition probabilities \mbox{$\bra{0}\Lambda_\beta (\ketbra{1}{1})\ket{0} = 1$}, \mbox{$\bra{1}\Lambda_\beta (\ketbra{0}{0})\ket{1} = e^{-\beta E}$}. 
This is the $\beta$-swap introduced in \cite{lostaglio2018elementary}, which can be realized by the dephasing thermalization
\begin{small} 
	\begin{align}\label{eq:betaswap}
	\Lambda_\beta(\rho_S)= & \sigma_- \rho_S \sigma_+ + e^{-\beta E} \sigma_+ \rho_S \sigma_- \\ \nonumber & + (1-e^{-\beta E})\sigma_- \sigma_+ \rho_S \sigma_- \sigma_+
	\end{align}
\end{small}
where $\sigma_+ = \ketbra{1}{0}$, $\sigma_- = \sigma_+^\dagger$. 
Since the unitary mapping the state to its correspondent maximally active state is the Pauli $X$ unitary, Theorem~\ref{th:dlevel} reads as follows:
\begin{restatable}{corollary}{thqubitprot}
	\label{th:qubitprot}
	Assume $E>0$, $\beta >0$. Without loss of generality (by making use of an initial diagonalizing unitary), we can take the initial state of the system to be diagonal in the energy basis with $p_0^{(0)} \geq p_1^{(0)}$. The optimal cooling protocol in $\mathcal{P}_\emptyset$ is such that in each round $k$:
	\begin{enumerate}
		\item The Pauli $X$ unitary is applied to $S$.
		\item The $\beta$-swap $\Lambda_\beta$ is applied to $S$.
	\end{enumerate}
	The population of the ground state after round $k$ is:
	\begin{equation}
	\label{eq:coolingqubit}
	p^{\left(k\right)}_0 = 1 - e^{-k\beta E}\left(1-p^{\left(0\right)}_0\right),
	\end{equation}
	and $p_0^{\left(k\right)}\rightarrow 1$ as $k\rightarrow \infty$.
\end{restatable}

The performance can be obtained by direct computation, or by setting $d=2$ in Eq.~\eqref{eq:ddimensional}. Note the simple structure of the optimal protocol, in particular the fact that the same operation is applied iteratively at each round. Furthermore, note that $\beta$-swaps cannot be realized by Markovian interactions with the thermal environment. In fact, if we optimized over Markovian interactions only, the optimal protocol with no auxiliary systems would be the trivial thermalization at the environment temperature $\rho_S \mapsto \tau_S$. Then, at best we would achieve a ground state population $1/(1+e^{-\beta E})$. The proof is given in the Methods (Sec.~\ref{ap:markov}). This shows that memory effects can be used to great advantage in cooling scenarios.  In particular, differently from standard HBAC protocols, our protocol achieves exponential convergence to the ground state. The `price' we pay is the need for greater control over the thermalization steps; however, we need no auxiliary systems in the unitary stage and the protocol iterates the same transformation at every round, which simplifies the implementation. \\

We presented an optimal protocol for qubits, but how is it realized by an explicit interaction and environment? Here we answer this question. It is known that an infinite dimensional environment is necessary to increase the purity of the target system to 1 in the absence of initial system-environment correlations \cite{ticozzi2014quantum, silva2016performance}; however, we do not need a complex environment: it was shown in \cite{lostaglio2018elementary} that a single bosonic mode in a thermal state
$\tau_B = \left(1-e^{-\beta E}\right)\sum_{n=0}^{\infty} e^{-n\beta E}\ketbra{n}{n}$ suffices to implement the required $\beta$-swap. Specifically, the unitary needed is
\begin{equation}
\label{eq:uSB}
U^{\beta}_{SB}=\ketbra{0,0}{0,0}+\sum_{n=1}^{\infty}\left(\ketbra{0,n}{1,n-1}+\ketbra{1,n-1}{0,n}\right).
\end{equation}
What is perhaps more remarkable is that, as we prove, we do not need to rethermalize (or otherwise refresh) the thermal mode at every round of the protocol. The \emph{same} mode can be reused in each round, in spite of the correlation build-up and the back-reaction, with no re-thermalisation needed. This provides a simplified and explicit version of Corollary~\ref{th:qubitprot} (proof in Methods, Sec.~\ref{ap:theorem2}): 

\begin{restatable}{theorem}{threusability} \label{th:reusability}
	Under the assumptions of Corollary~\ref{th:qubitprot}, the optimal protocol in $\mathcal{P}_{\emptyset}$ has the initial state $\rho^{\left(0\right)}_S \otimes\tau_B$ and is such that in each round $k$:
	\begin{enumerate}
		\item The Pauli $X \otimes \mathbb{I} $ unitary is applied to $\rho^{\left(k-1\right)}_{SB}$, the state of system-bath after round $k-1$.
		\item The unitary $U^\beta_{SB}$ is applied to $X\otimes \mathbb{I} ( \rho^{\left(k-1\right)}_{SB}) X \otimes \mathbb{I}$.
	\end{enumerate}
\end{restatable}

\subsection{Robustness to imperfections}

Theorem~\ref{th:reusability} holds even beyond some of the (standard) idealizations we made:
\begin{enumerate}
	\item If the bosonic mode gap does not perfectly match the system gap, $U^\beta_{SB}$ still realises the cooling of Eq.~\eqref{eq:coolingqubit}, with the caveat that some work flows at each $\beta$-swap.
	\item We can consider a typical imperfection, i.e. a small anharmonicity in the ladder of $B$. In the Methods (Sec.~\ref{ap:anharmonicity}) we show that taking $E_{n+1}- E_n = E(1-(n+1)\tau^2) + o(\tau^2)$, the protocol of Theorem~\ref{th:reusability} performs very closely to the ideal scenario of Eq.~\eqref{eq:coolingqubit} even for moderate anharmonicity (within $0.005 \%$ for $\beta E$=1, $\tau = 0.05$).
\end{enumerate}

Furthermore, up to now we considered an idealized scenario in which one can perform the required $\beta$-swap exactly. In a real experiment, however, one will only realize an approximation of it (we will see an explicit model in the next section). Consider a noise model in which, instead of the de-excitation probability $1$ required by the $\beta$-swap, one can only realize dephasing thermalizations $\Lambda^{(k)}$ with an induced de-excitation probability $\bra{0} \Lambda^{(k)}(\ketbra{1}{1})\ket{0} \leq 1-\epsilon$. Denote this set by $\mathcal{P}^\epsilon_\emptyset$. Define an $\epsilon$-\emph{noisy $\beta$-swap} as the dephasing thermalization $\Lambda^\epsilon_{\beta}$ with $\bra{0} \Lambda^\epsilon_{\beta}(\ketbra{1}{1})\ket{0} = 1-\epsilon$ 
(when $\epsilon=0$ this is the $\beta$-swap). Then one can derive the following noise-robust version of Theorem~\ref{th:qubitprot}:  

\begin{restatable}{theorem}{thmrobustness} \label{thm:robustness}
	Under the assumptions of Corollary~\ref{th:qubitprot} and given $\epsilon \leq \frac{1}{1+e^{\beta E} + e^{2\beta E}} $, the optimal nontrivial cooling protocol in $\mathcal{P}^\epsilon_\emptyset$ is such that in each round $k$:
	\begin{enumerate}
		\item The Pauli $X$ unitary is applied to $S$.
		
		\item The $\epsilon$-noisy $\beta$-swap $\Lambda^\epsilon_{\beta}$ is applied to $S$.
	\end{enumerate}
	The population of the ground state after round $k$ is:
	\begin{align}
	\label{eq:coolingrobustness}
	p_{\rm 0}^{(k)} =&1-\frac{\epsilon}{2-(1-\epsilon) Z} \\
	&\,\,- \left((1-\epsilon) Z -1\right)^k \left( 1-\frac{\epsilon}{2-(1-\epsilon) Z}- p_0^{(0)}\right),\nonumber
	\end{align}
	where $Z = 1 + e^{-\beta E}$ and $p^{\left( k\right)}_0\rightarrow 1-\frac{\epsilon}{2-(1-\epsilon)Z}$ as $k\rightarrow\infty$.
\end{restatable}

In words: even in the presence of (moderate) noise, the optimal strategy is to perform the best approximation to the ideal protocol; hence,  the simple structure of the optimal protocols is robust to imperfections. Theorem~\ref{thm:robustness} is proved in Section \ref{ap:opt qubit} by a tedious but straightforward optimization. There, we also prove another form of `robustness': if we optimize over the \emph{larger} set of thermalization models known as \emph{thermal operations} \cite{brandao2011resource}, which include protocols where some amount of superposition in the energy basis survives the thermalization step, Theorem~\ref{thm:robustness} holds unchanged. Surprisingly, the extra coherent control in the thermalization phase is not necessary for optimal cooling.

\subsection{An experimental proposal}

We can now provide an explicit experimental proposal. More details about the calculations and simulations involved can be found in Sec.~\ref{ap:JC} -\ref{ap:JCreuse}.

The unitary $U^\beta_{SB}$ of Eq.~\eqref{eq:uSB} can be realized exactly with an intensity-dependent Jaynes-Cummings model \cite{naderi2005theoretical,aberg2014catalytic}, $\tilde{H}_{\textrm{JC}} = g( \sigma_+ \otimes (a a^\dag)^{-1/2} a + \sigma_- \otimes (a a^\dag)^{-1/2} a^\dag)$. However, a perhaps more promising avenue is to approximate the $\beta$-swap steps with a resonant Jaynes-Cummings (JC) coupling with a thermal bosonic mode, \mbox{$H_{\textrm{JC}} = g( \sigma_+ \otimes a + \sigma_- \otimes a^\dag)$}. Assuming good control of the interaction time $s$, we can numerically optimize $s$ to realize an $\epsilon$-noisy $\beta$-swap with the small $\epsilon$.  If the thermal mode is reset at each round, we can compute the performance of this implementation using Theorem~\ref{thm:robustness}.

\begin{figure}[t!]
	\includegraphics[width=1.03\linewidth]{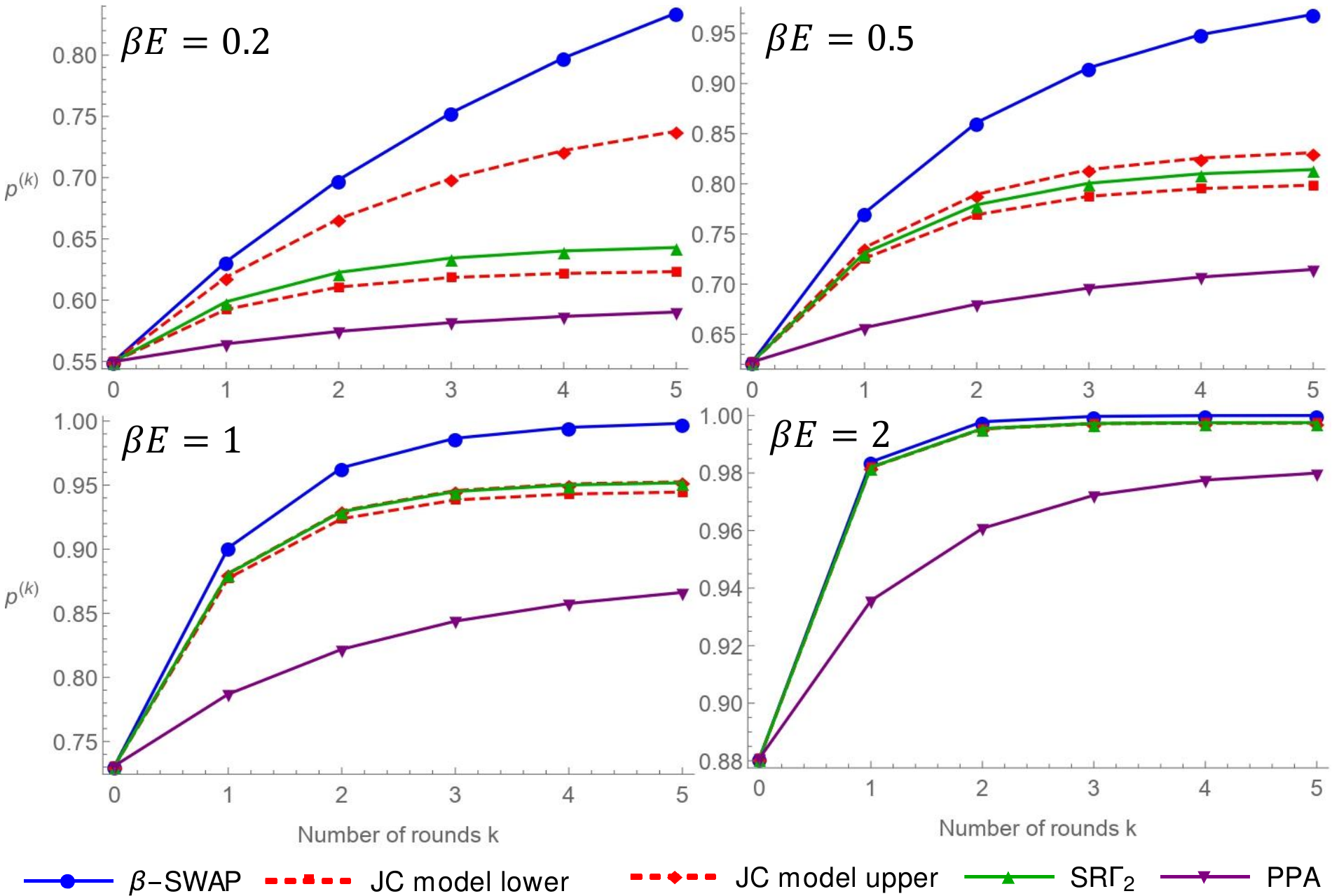}
	\caption{\small Ground state occupation $p^{(k)}_0$ at step $k$ for different environment temperatures for: ideal optimal cooling protocol from Theorem~\ref{th:qubitprot} (blue), upper and lower bounds on a JC realization of this protocol (red, see Sec.~\ref{ap:JC} for details on how to calculate them), SR$\Gamma$ protocol from \cite{rodriguez2017heat} run with 1 ancilla qubit (green), PPA protocol from \cite{schulman2005physical} run with 2 ancilla qubits (purple). The initial state for all protocols is thermal.}
	\label{fig:comparison}
\end{figure}

In Fig.~\ref{fig:comparison} we compare this protocol with leading proposals in the literature, for an initially thermal target. The JC protocol
outperforms the PPA with 2 ancillas \cite{schulman2005physical} (even with some limitations in the timing accuracy and the total available waiting time), with the exclusion of the very high temperature regime; when $\beta E$ is not too small, it performs comparably (if potentially slightly worse) to the ideal version of the non-local thermalization scheme with 1 ancilla proposed in \cite{rodriguez2017heat}. The optimal $\beta$-swap protocol outperforms all these protocols, but requires an intensity-dependent JC model whose implementation we leave as an open question. 

\begin{figure}[t!]
	\centering
	\includegraphics[width=0.8\linewidth]{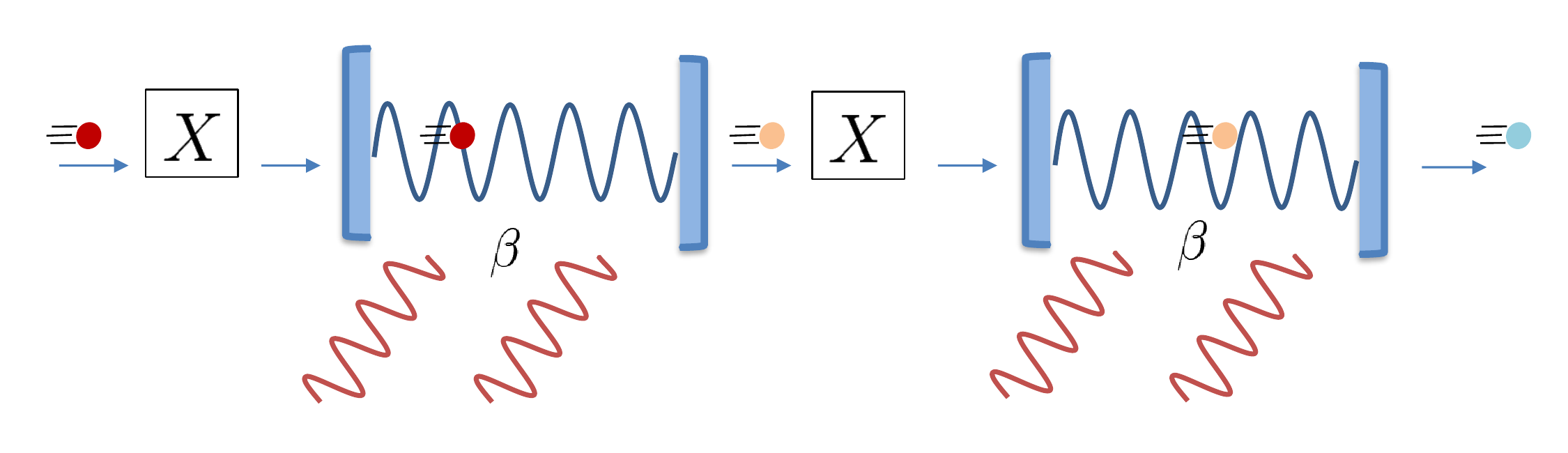}
	\caption{\small The optimal protocol for cooling can be approximated by one in which a population inversion (Pauli $X$) is applied to the qubit before it enters the cavity, where it interacts with a resonant mode initially at temperature $\beta$. If the interaction parameters are appropriately chosen, such that a small $\epsilon$ is achieved, the outgoing qubit has a much lower temperature.}
	\label{fig:cavities}
\end{figure}

\begin{figure}[t!]
	\centering
	\includegraphics[width=
	0.7\linewidth]{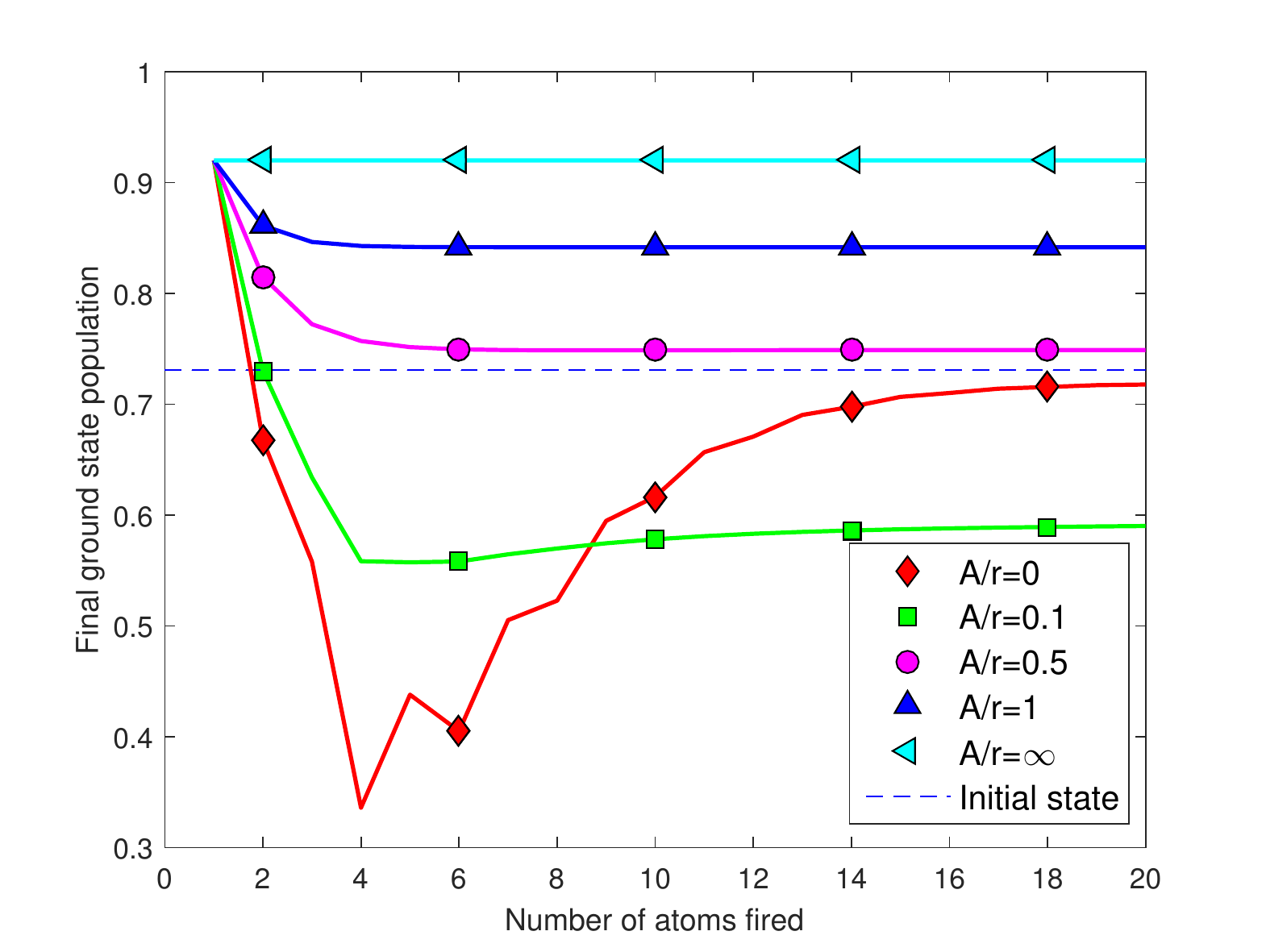}
	\caption{Cooling achieved on each atom by two rounds of the Pauli/Jaynes-Cummings protocols, as a function of the number of atoms already fired. Each curve represents a different ratio between the re-thermalization rate $A$ and the rate $r$ at which the atoms are fired. The atoms are prepared in an initially thermal state with $\beta E =1$, the coupling strength is $g=1$ and time of interaction $t=98.92$.}
	\label{fig:JC2cavity}
\end{figure}

These considerations assume that at every round the bosonic mode is reset to the thermal state. However we show, using a standard master equation (Eq.~\eqref{eq:masterequationcavity} in Sec.~\ref{ap:JCreuse}), that the reset can be substituted by a more realistic slow rethermalization of the single mode with an external environment. Since reasonably high cooling is achieved after 2 rounds, this suggests the following implementation: a stream of slowly fired atoms passes through two identical cavities resonant with the qubits we are trying to cool, each supporting a single mode initially thermal; at the entrance of each cavity we perform a Pauli $X$ operation, and let each qubit interact with the cavity mode for a chosen time (see Fig.~\ref{fig:cavities} for a schematic description). Numerics show that if re-thermalization of the cavity mode is sufficiently quick compared to the firing rate, the cooling performance settles to a constant as many atoms are cooled (see Fig.~\ref{fig:JC2cavity}). 

This is, to our knowledge, the most appealing setting to experimentally implement the protocol and is, in fact, highly reminiscent of a micromaser \cite{filipowicz1986theory, walther2006cavity}. This device consists of a cavity with a harmonic oscillator in an initially thermal state, and it works provided we are able to keep this oscillator out of thermal equilibrium. The firing of an excited atom through a cavity in the micromaser can be seen as a single instance of our proposed imperfect implementation of the optimal cooling protocol. However, the figures of merit in each case are different: in the micromaser, we need very pure atoms in order to excite the cavity efficiently, while in the cooling protocol the aim is to obtain these very pure atoms in the first place. Nevertheless, this suggests that experimental settings where the micromaser has been shown to be possible might be good platforms in which to test our algorithm. To our knowledge, this currently includes both cavity QED \cite{filipowicz1986theory,walther2006cavity} and solid-state settings \cite{rodrigues2007quantum}.

In conclusion, often HBAC techniques made the implicit assumption that the best way of exploiting the external environment to pump entropy away from the system is to thermalize the auxiliary ancillas. Here we show how optimizing over the thermalization strategy provides new cooling protocols breaking previously established limits, in particular allowing exponential convergence to the ground state in the ideal scenario.

\widetext

\section{Methods}

\subsection{$\beta$-permutations and the thermal polytope}
\label{ap:thermalpolytope}

Here we discuss how to construct $\beta$-permutations and how they characterize the thermal polytope. The first key definition is a generalization of the concept of majorization. Recall that, given two $d-1$-dimensional probabilities $\v{p}$ and $\v{p}'$, $\v{p}$ majorizes $\v{p}'$, denoted $\v{p} \succeq \v{p}'$, if $\sum_{i=0}^k p_i^\downarrow \geq \sum_{i=0}^k {p'}_i^\downarrow$ for $k=0,...,d-2$, where $\v{x}^\downarrow$ denotes the vector $\v{x}$ arranged in descending order. $\rho \succeq \rho'$ is defined as the same relation among the corresponding eigenvalues. Then define 
\begin{definition}[Thermo-majorization \cite{horodecki2013fundamental}]
	Given a state $\rho$ with Hamiltonian $H=\sum_{i=0}^{d-1}E_i\ketbra{i}{i}$, let $\v{p}=\left(p_0,\dots,p_{d-1}\right)$ where $p_i=\bra{i}\rho \ket{i}$ and $\v{E}=\left(E_1,\dots,E_d\right)$. The thermo-majorization curve of $\rho$ is formed by:
	\begin{enumerate}
		\item Applying a permutation $\pi$ of $\{0,...,d-1\}$ to both $\v{p}$ and $\v{E}$ such that $p_{\pi(i)} e^{\beta E_{\pi(i)}}$ is in non-increasing order. We refer to the vector $\pi$ with elements $\pi(i)$ as the \emph{$\beta$-order} of~$\v{p}$.
		
		\item Plotting the ordered `elbow' points $(0,0)$, \mbox{$\left\{\left(\sum_{i=0}^{k}e^{-\beta E_{\pi(i)}},\sum_{i=0}^{k} p_{\pi(i)}\right)\right\}_{k=0}^{d-1}$} and connecting them piecewise linearly to form a concave curve - the thermo-majorization curve of $\v{p}$.
	\end{enumerate}
	Given two probability distributions $\v{p}$ and $\v{p}'$ associated with the same energy levels, we say that $\v{p}$ thermo-majorizes $\v{p}'$ if the thermo-majorization curve of $\v{p}$ is never below that of $\v{p}'$. We denote this by $\v{p} \succeq_{\textrm{th}} \v{p}'$. Also, if $\rho$ and $\rho'$ are states with population vectors $\v{p}$ and $\v{p}'$, the notation $\rho \succeq_{\textrm{th}} \rho'$ denotes $\v{p} \succeq_{\textrm{th}} \v{p}'$.  
\end{definition}
Note that $\succeq_{th}$ becomes $\succeq$ in the infinite temperature limit $\beta \rightarrow 0$. One can show that $\v{p} \succeq_{\textrm{th}} \v{p}'$ is equivalent to the existence of a Gibbs stochastic matrix (a stochastic matrix $G$ with $G \v{g} = \v{g}$, if $g_i \propto e^{-\beta E_i}$) such that $G\v{p}=\v{p}'$~\cite{ruch1980generalization}. Furthermore, a dephasing thermalization $\Lambda$ acts on the population vector as a Gibbs-stochastic matrix $G_{j|i} = \bra{j}\Lambda(\ketbra{i}{i})\ket{j}$;  conversely, every Gibbs-stochastic matrix can be realized by dephasing thermalizations. As such, the thermal polytope of $\v{p}$ coincides with the set of probability distributions $\v{p}'$ such that $\v{p} \succeq_{\textrm{th}} \v{p}'$. That this is a convex polytope follows from the fact that the set of Gibbs-stochastic matrices also is. To characterize the thermal polytope of $\v{p}$ explicitly, we construct a set of maps, called $\beta$-permutations, that take $\v{p}$ to each one of its extremal points. 

\emph{Algorithm to construct $\Lambda^{(\pi,\alpha)}$.} By the discussion above each $\beta$-permutation $\Lambda^{(\pi,\alpha)}$ is fully characterized by a matrix of transition probabilities $P^{(\pi,\alpha)}_{j|i} = \bra{j}\Lambda^{(\pi,\alpha)}(\ketbra{i}{i})\ket{j}$. Let $\pi$ and $\alpha$ each be one of the $d! -1$ possible $\beta$-orders. Order the rows of $P^{(\pi,\alpha)}$ according to $\alpha$ and the columns according to $\pi$:
\begin{equation*}
G =
\begin{pmatrix}
P^{(\pi,\alpha)}_{\alpha(0) \pi(0)} & \dots & P^{(\pi,\alpha)}_{\alpha(0)\pi(d-1)} \\
\vdots & & \vdots \\
P^{(\pi,\alpha)}_{\alpha(d-1) \pi(0)} & \dots & P^{(\pi,\alpha)}_{\alpha(d-1)\pi(d-1)} 
\end{pmatrix}
.
\end{equation*}
Then $G$ is constructed as follows.

\emph{Row 0}: If $e^{-\beta E_{\alpha\left(0\right)}}< e^{-\beta E_{\pi\left(0\right)}}$, set:
\begin{align*}
k_0=1, \;\; 	G_{00}=\frac{e^{-\beta E_{\alpha\left(0\right)}}}{e^{-\beta E_{\pi\left(0\right)}}} \label{eq:G_11a}, \;\; G_{0j}=0,\quad j \in [1,d-1],
\end{align*}
else, let $k_0$ be the smallest integer such that:
\begin{equation*}
\sum_{j=0}^{k_0}e^{-\beta E_{\pi\left(j\right)}} \geq e^{-\beta E_{\alpha\left(0\right)}}. \label{eq:k1}
\end{equation*}
and set:
\begin{align*}
G_{0j}&=1,\quad 0\leq j\leq k_0-1\\
G_{0k_0} &=\frac{e^{-\beta E_{\alpha\left(0\right)}}-\sum_{j=0}^{k_0-1}e^{-\beta E_{\pi\left(j\right)}}}{e^{-\beta E_{\pi\left(k_0\right)}}} \label{eq:G_1k_1}\\
G_{0j}&=0,\quad j \in [k_0+1, d-1]
\end{align*}

\emph{Row $m\geq 1$}: If $\sum_{i=0}^{m}e^{-\beta E_{\alpha\left(i\right)}}<\sum_{j=0}^{k_{m-1}}e^{-\beta E_{\pi\left(j\right)}}$, set:
\begin{small}
	\begin{align*}
	k_m&=k_{m-1}, \; \; 	G_{mj}=0,\quad 0\leq j\leq k_{m-1}-1 \\
	G_{mk_{m-1}}&=\frac{e^{-\beta E_{\alpha\left(m\right)}}}{e^{-\beta E_{\pi\left(k_{m-1}\right)}}}, \; 	G_{mj}=0,\; j \in[k_{m-1}+1,d-1]
	\end{align*}
\end{small}
else, let $k_m$ be the smallest integer such that:
\begin{equation}
\sum_{j=0}^{k_m} e^{-\beta E_{\pi\left(j\right)}} \geq \sum_{i=0}^{m} e^{-\beta E_{\alpha\left(i\right)}}.
\end{equation}
and set:
\begin{align*}
G_{mj}=0,\; j& \in [0,k_{m-1}-1], \;
G_{m k_{m-1}}=1-\sum_{i=0}^{m-1} G_{i k_{m-1}} \\
G_{m j}&=1,\quad  j \in[k_{m-1}+1, k_m -1] \\
G_{m k_m} &= \frac{\sum_{i=0}^{m} e^{-\beta E_{\alpha\left(i\right)}}-\sum_{j=0}^{k_m-1}e^{-\beta E_{\pi\left(j\right)}}}{e^{-\beta E_{\pi\left(k_m\right)}}}\\
G_{m j} &=0,\quad j \in [k_{m}+1, d-1].
\end{align*}

In the next section, we prove the important fact that, if $\pi$ is the $\beta$-order of $\v{p}$, the set $\{P^{(\pi,\alpha)}\v{p}\}_\alpha$, for $\alpha$ varying over all possible $\beta$-orders, includes all extremal points of the thermal polytope of~$\v{p}$ (Lemma~\ref{le:maxbeta}). 

\subsection{Proof of properties of $\beta$-permutations}

In the following we show that the $\beta$-permutations, defined Section \ref{sec:general}, convert $\v{p}$ into any of the extremal points of the thermal polytope $\v{p}^\alpha$, while being independent of the specific form of $\v{p}$ (the set of $\beta$-permutation that need to be applied only depends on the $\beta$-order of $\v{p}$). This is in complete analogy with permutations and their role within the theory of doubly-stochastic matrices as determined by Birkhoff's theorem \cite{birkhoff1946tres} (in fact, $\beta$-permutations become permutations in the infinite temperature limit).

Specifically, we will prove the following facts:
\begin{enumerate}
	\item $P^{(\pi,\alpha)}$ is a Gibbs-stochastic map.
	\item Given $\v{p}$ with $\beta$-order $\pi$, $\v{p}^\alpha := P^{(\pi,\alpha)} \v{p}$ has $\beta$-order $\alpha$ and is such that there is no $\v{q}$ with $\beta$-order $\alpha$ such that $\v{p} \succeq_{th} \v{q} \succeq_{th} \v{p}^\alpha$ (Lemma~\ref{le:maxbeta}).
	\item The set of $\v{p}^\alpha$ for varying $\alpha$ includes all the extremal points of the themal polytope of $\v{p}$ (the latter is equivalently defined as all $\v{q}$ such that $\v{p} \succeq_{th} \v{q}$). 
\end{enumerate}

\subsubsection{Proof of claim 1}
Recall that we reordered the rows of $P^{(\pi,\alpha)}$ according to $\alpha$ and the columns according to $\pi$:
\begin{equation*}
G =
\begin{pmatrix}
P^{(\pi,\alpha)}_{\alpha(0) \pi(0)} & \dots & P^{(\pi,\alpha)}_{\alpha(0)\pi(d-1)} \\
\vdots & & \vdots \\
P^{(\pi,\alpha)}_{\alpha(d-1) \pi(0)} & \dots & P^{(\pi,\alpha)}_{\alpha(d-1)\pi(d-1)} 
\end{pmatrix}
.
\end{equation*}
Hence, the condition of Gibbs-stochasticity can be rewritten as
\begin{align}
G_{ij}&\geq0 \label{eq:pos}\\
\sum_{i=0}^{d-1} G_{ij}&=1 \label{eq:stoc}\\
\sum_{j=0}^{d-1} G_{ij}e^{-\beta E_{\pi\left(j\right)}}&=e^{-\beta E_{\alpha\left(i\right)}} \label{eq:gibbs pres}.
\end{align}
For simplicity, we repeat here the algorithm for constructing $P^{(\pi,\alpha)}$. Row 0 is populated as follows:
\begin{algorithmic}
	\If{$e^{-\beta E_{\alpha\left(0\right)}}< e^{-\beta E_{\pi\left(0\right)}}$}
	\State{Set:\begin{align}
		k_0&=1\\
		G_{00}&=\frac{e^{-\beta E_{\alpha\left(0\right)}}}{e^{-\beta E_{\pi\left(0\right)}}} \label{eq:G_11a}\\
		G_{0j}&=0,\quad 1\leq j\leq d-1
		\end{align}}
	\Else
	\State{Let $k_0$ be the smallest integer such that:
		\begin{equation}
		\sum_{j=0}^{k_0}e^{-\beta E_{\pi\left(j\right)}} \geq e^{-\beta E_{\alpha\left(0\right)}}. \label{eq:k1}
		\end{equation}}
	\State{Set:
		\begin{align}
		G_{0j}&=1,\quad 0\leq j\leq k_0-1\\
		 G_{0k_0}&=\frac{e^{-\beta E_{\alpha\left(0\right)}}-\sum_{j=0}^{k_0-1}e^{-\beta E_{\pi\left(j\right)}}}{e^{-\beta E_{\pi\left(k_0\right)}}} \label{eq:G_1k_1}\\
		 G_{0j}&=0,\quad k_0+1\leq j\leq d-1
		\end{align}
	}
	\EndIf
\end{algorithmic}

\begin{figure}[t!]
	\centering
	\includegraphics[width=0.3\linewidth]{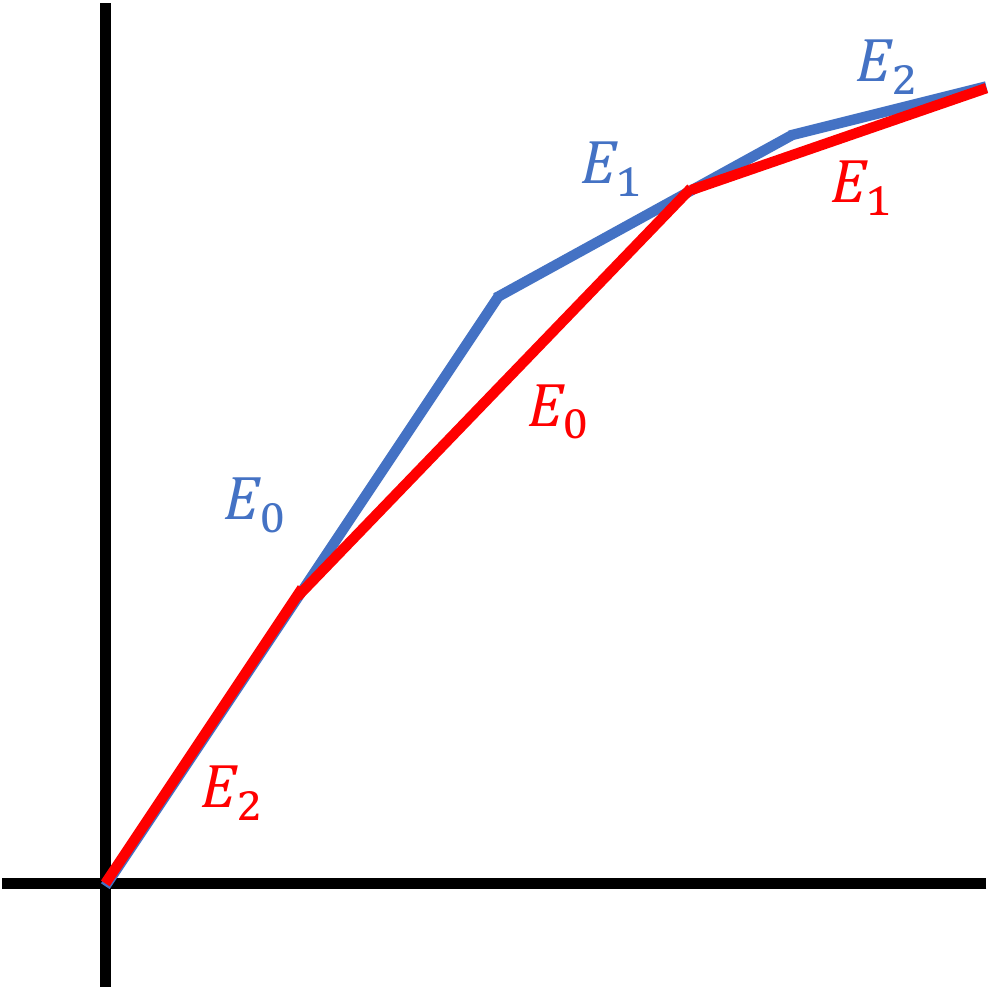}
	\caption{Here we illustrate the notion of a \emph{maximal $\beta$-order state}. The state $\v{p}$, shown in blue, has $\beta$-order $\pi=\left(0,1,2\right)$. The state $\v{p}^{\alpha}$ for $\alpha=\left(2,0,1\right)$ is shown in red and has a simple geometrical interpretation. The choice of $\alpha$ (Condition~\ref{max_beta_condition1} in Sec. ~\ref{sec:claim2}) fixes the $x$-axis points to be $x_0 = e^{-\beta E_2}$, $x_1 = e^{-\beta E_2} +  e^{-\beta E_0}$, $x_2 = e^{-\beta E_2} +  e^{-\beta E_0} + e^{-\beta E_1}$. Then Condition~\ref{max_beta_condition2} is equivalent to the request that the curve in red touches the blue curve at these points.}
	\label{fig:max beta}
\end{figure}

Let us first check that Eqs.~\eqref{eq:pos} and \eqref{eq:gibbs pres} are fulfilled in each clause of row $m=0$. In the first clause, it is clear that $G_{00}\geq0$.
Gibbs preservation follows as:
\begin{equation*}
\sum_{j=0}^{d-1} G_{0j}e^{-\beta E_{\pi\left(j\right)}}=G_{00}e^{-\beta E_{\pi\left(0\right)}}=e^{-\beta E_{\alpha\left(0\right)}}.
\end{equation*}

For the second clause, it is again true that we have $0\leq G_{0j} \leq 1$, for all $j$ (if $G_{0k_0}$ were greater than $1$, then this would contradict the definition of $k_0$ in Eq.~\eqref{eq:k1}). 
We then note that:
\begin{align*}
\sum_{j=0}^{d-1} G_{0j}e^{-\beta E_{\pi\left(j\right)}}=\sum_{j=0}^{k_0-1}e^{-\beta E_{\pi\left(j\right)}}+e^{-\beta E_{\alpha\left(0\right)}} - \sum_{j=0}^{k_0-1}e^{-\beta E_{\pi\left(j\right)}}=e^{-\beta E_{\alpha\left(0\right)}}
\end{align*}
so Gibbs preservation is satisfied. 

With row zero in place, row $m$ ($m\in\left\{1,\dots d-1\right\}$) is populated as follows:
\begin{algorithmic}
	\If{$\sum_{i=0}^{m}e^{-\beta E_{\alpha\left(i\right)}}<\sum_{j=0}^{k_{m-1}}e^{-\beta E_{\pi\left(j\right)}}$}
	\State{Set:
		\begin{align}
		k_m&=k_{m-1}\\
		G_{mj}&=0,\quad 0\leq j\leq k_{m-1}-1\\
		G_{mk_{m-1}}&=\frac{e^{-\beta E_{\alpha\left(m\right)}}}{e^{-\beta E_{\pi\left(k_{m-1}\right)}}} \label{eq:G if}\\
		G_{mj}&=0,\quad k_{m-1}+1\leq j \leq d-1
		\end{align}
	}
	\Else
	\State{Let $k_m$ be the smallest integer such that:
		\begin{equation}
		\sum_{j=0}^{k_m} e^{-\beta E_{\pi\left(j\right)}} \geq \sum_{i=0}^{m} e^{-\beta E_{\alpha\left(i\right)}}
		\end{equation}
	}
	\State{Set:
		\begin{align}
		G_{mj}&=0,\quad 0\leq j\leq k_{m-1}-1 \label{eq:m stoc1}\\
		G_{m k_{m-1}}&=1-\sum_{i=0}^{m-1} G_{i k_{m-1}} \label{eq:m stoc2}\\
		G_{m j}&=1,\quad k_{m-1}+1 \leq j \leq k_m -1 \label{eq:m stoc3}\\
		G_{m k_m} &= \frac{\sum_{i=0}^{m} e^{-\beta E_{\alpha\left(i\right)}}-\sum_{j=0}^{k_m-1}e^{-\beta E_{\pi\left(j\right)}}}{e^{-\beta E_{\pi\left(k_m\right)}}} \label{eq:mk_m}\\
		G_{m j} &=0,\quad k_{m}+1 \leq j \leq d-1
		\end{align}
	}
	\EndIf
\end{algorithmic}

We now check that Eq.~\eqref{eq:pos} and Eq.~\eqref{eq:gibbs pres} are fulfilled in each of the clauses for each row in the matrix. 
For the first clause $0\leq G_{mk_{m-1}}\leq 1$ 
follows from:
\begin{align*}
e^{-\beta E_{\alpha(m)}} &\leq \sum_{i=0}^{k_{m-1}} e^{-\beta E_{\pi(i)}} - \sum_{i=0}^{m-1} e^{-\beta E_{\alpha(i)}} \\&= e^{-\beta E_{\pi(k_{m-1})}} + \sum_{i=0}^{k_{m-1}-1} e^{-\beta E_{\pi(i)}} - \sum_{i=0}^{m-1} e^{-\beta E_{\alpha(i)}} \\& \leq e^{-\beta E_{\pi(k_{m-1})}},
\end{align*}
where in the last inequality we used the definition of $k_{m-1}$.
Finally, note that:
\begin{equation*}
\sum_{j=0}^{d-1} G_{m j}e^{-\beta E_{\pi\left(j\right)}}=G_{m k_{m-1}}e^{-\beta E_{\pi\left(k_{m-1}\right)}}=e^{-\beta E_{\alpha\left(m\right)}}
\end{equation*}
and hence Eq.~\eqref{eq:gibbs pres} holds.

For the second clause,	the definitions of $k_{m-1}$ and $k_m$ ensure that both $0\leq G_{m k_{m-1}}\leq 1$ and $0\leq G_{m k_m}\leq 1$ hold. 
We now show Gibbs preservation. First note that:
\begin{align}
\sum_{j=0}^{d-1} G_{m j}e^{-\beta E_{\pi\left(j\right)}}&=G_{m k_{m-1}} e^{-\beta E_{\pi\left(k_{m-1}\right)}}+\sum_{j=k_{m-1}+1}^{k_m-1}e^{-\beta E_{\pi\left(j\right)}}+G_{m k_m}e^{-\beta E_{\pi\left(k_m\right)}} \nonumber\\
&=\left(1-\sum_{i=0}^{m-1} G_{i k_{m-1}}\right) e^{-\beta E_{\pi\left(k_{m-1}\right)}} +\sum_{j=k_{m-1}+1}^{k_m-1}e^{-\beta E_{\pi\left(j\right)}} +\sum_{i=0}^{m} e^{-\beta E_{\alpha\left(i\right)}}-\sum_{j=0}^{k_m-1} e^{-\beta E_{\pi\left(j\right)}} \label{eq:dagger}
\end{align} 
where we have used Eq.~\eqref{eq:m stoc2} and Eq.~\eqref{eq:mk_m} in the last line. Consider now $\sum_{i=0}^{m-1} G_{i k_{m-1}}$. Note that there exists an integer $r$ such that $0\leq r\leq m-1$ and:
\begin{align*}
G_{i k_{m-1}}&=0,\quad i<r\\
G_{r k_{m-1}}&=\frac{\sum_{i=0}^{r}e^{-\beta E_{\alpha\left(i\right)}}-\sum_{j=0}^{k_r-1}e^{-\beta E_{\pi\left(j\right)}}}{e^{-\beta E_{\pi\left(k_r\right)}}}\\
G_{i k_{m-1}}&=\frac{e^{-\beta E_{\alpha\left(i\right)}}}{e^{-\beta E_{\pi\left(k_i\right)}}},\quad r+1\leq i\leq m-1
\end{align*}
with $k_i=k_r$ for $r\leq i \leq m-1$. Using these expressions in Eq.~\eqref{eq:dagger}, we see that:
\begin{align*}
\sum_{j=0}^{d-1} G_{m j}e^{-\beta E_{\pi\left(j\right)}}&=\left(1-\frac{\sum_{i=0}^{m-1}e^{-\beta E_{\alpha\left(i\right)}}-\sum_{j=0}^{k_{m-1}-1}e^{-\beta E_{\pi\left(j\right)}}}{e^{-\beta E_{\pi\left(k_{m-1}\right)}}}\right)e^{-\beta E_{\pi\left(k_{m-1}\right)}}\\
&\quad\quad+\sum_{j=k_{m-1}+1}^{k_m-1}e^{-\beta E_{\pi\left(j\right)}} +\sum_{i=0}^{m} e^{-\beta E_{\alpha\left(i\right)}}-\sum_{j=0}^{k_m-1} e^{-\beta E_{\pi\left(j\right)}}\\
&=\sum_{j=0}^{k_m-1}e^{-\beta E_{\pi\left(j\right)}} + e^{-\beta E_{\alpha\left(m\right)}} - \sum_{j=0}^{k_m-1}e^{-\beta E_{\pi\left(j\right)}}\\
&= e^{-\beta E_{\alpha\left(m\right)}}
\end{align*}
and hence the Gibbs distribution is preserved.

Finally, let us show that the matrix constructed is stochastic. Using the expressions for $G_{i k_{m-1}}$ given above, the fact that $k_i = k_r$ for all $r \leq i \leq m-1$ and the definition of $k_{m-1}$,
\begin{equation*}
\sum_{i=0}^{m-1} G_{i k_{m-1}} = \frac{\sum_{i=0}^{m-1} e^{-\beta E_{\alpha(i)}} - \sum_{i=0}^{k_{m-1}-1} e^{-\beta E_{\pi(i)} }}{e^{-\beta E_{\pi(k_{m-1})}}}<1.
\end{equation*}
Together with Eq.~\eqref{eq:m stoc2} and the already proved fact that $G_{ij} \in [0,1]$ for all $i,j \in\left\{0,\dots,d-1\right\}$, we get that $G$ is stochastic.

\subsubsection{Proof of claim 2}\label{sec:claim2}

\begin{figure}[t]
	\begin{minipage}{.5\columnwidth}
		\centering
		\includegraphics[width=0.6\textwidth]{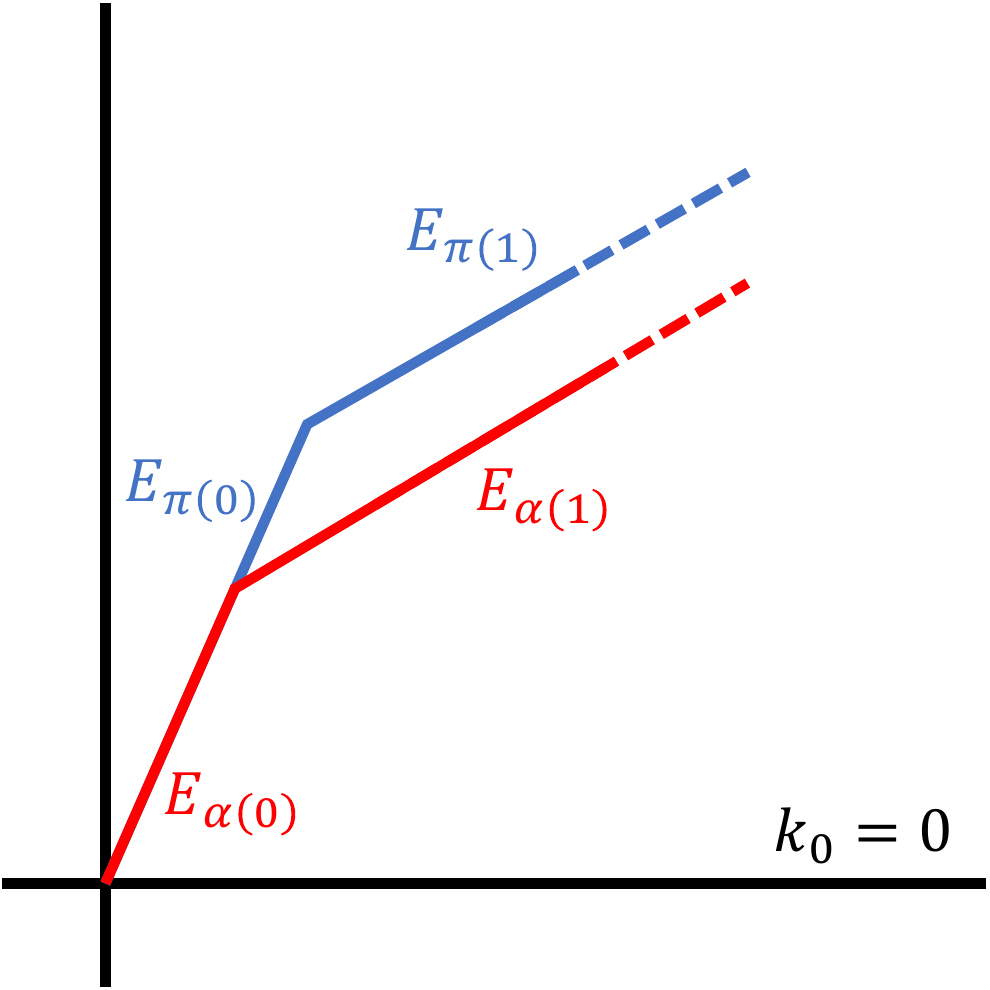}
		\label{fig:row1 case1}
		\centerline{(a) Row 0, case 1}
	\end{minipage}
	\begin{minipage}{.5\columnwidth}
		\centering
		\includegraphics[width=.6\textwidth]{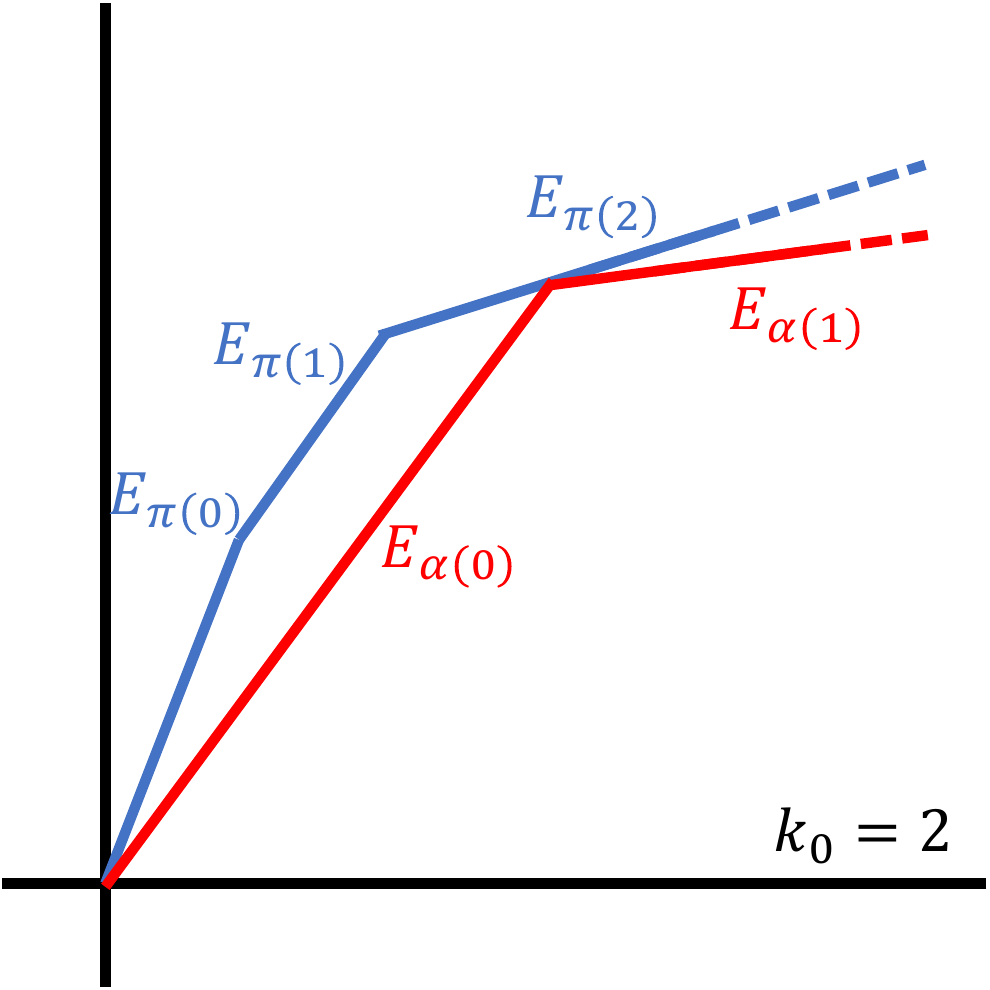}
		\label{fig:row1 case2}
		\centerline{(b) Row 0, case 2}
	\end{minipage}
	\caption{Thermo-majorization diagrams illustrating the construction of row zero of $P^{(\pi, \alpha)}$. In case 1, \mbox{$e^{-\beta E_{\alpha\left(0\right)}}< e^{-\beta E_{\pi\left(0\right)}}$} holds, while in case 2 it does not.}
	\label{fig:row1}
\end{figure}

This can be stated as the following Lemma:
\begin{lemma}
	\label{le:maxbeta}
	If $\v{p}$ has $\beta$-order $\pi$ and $\v{p}^\alpha:= P^{(\pi,\alpha)}\v{p}$, then $\v{p}^\alpha$ has $\beta$-order $\alpha$ and is `maximal' in the sense that there is no $\v{q}$ with $\beta$-order $\alpha$ such that $
	\v{p} \succeq_{\textrm{th}} \v{q} \succeq_{\textrm{th}} \v{p}^{\alpha}.$
\end{lemma}
We now construct the proof.
$\beta$-permutations have a simple geometrical description in terms of thermo-majorization curves. In row zero, we want to maximize the population of $E_{\alpha\left(0\right)}$ subject to the thermo-majorization constraints. To do this, we compare $e^{-\beta E_{\alpha\left(0\right)}}$ and  $e^{-\beta E_{\pi\left(0\right)}}$. If $e^{-\beta E_{\alpha\left(0\right)}}<e^{-\beta E_{\pi\left(0\right)}}$, then we cannot move all of the population in $E_{\pi\left(0\right)}$ to $E_{\alpha\left(0\right)}$ without violating thermo-majorization and must instead move only a fraction of it, as given by Eq.~\eqref{eq:G_11a}. This case is illustrated in Fig.~\ref{fig:row1}a. On the other hand, if $e^{-\beta E_{\alpha\left(0\right)}}\geq e^{-\beta E_{\pi\left(0\right)}}$, then we can move all of the population in $E_{\pi\left(0\right)}$ to $E_{\alpha\left(0\right)}$ and set $G_{00}=1$. We then try to move population from $E_{\pi\left(1\right)}$ to $E_{\alpha\left(0\right)}$ and repeat this process until we reach an energy level whose population we cannot move into $E_{\alpha\left(0\right)}$ without violating thermo-majorization. This energy level is defined through Eq.~\eqref{eq:k1} and given by $E_{\pi\left(k_0\right)}$. From $E_{\pi\left(k_0\right)}$ we can only move a fraction of the population into $E_{\alpha\left(0\right)}$, given by Eq.~\eqref{eq:G_1k_1}. When we reach this point, we have moved as much population as possible into  $E_{\alpha\left(0\right)}$. This case is illustrated in Fig.~\ref{fig:row1}b.

\begin{figure}[t]
	\begin{minipage}{.5\columnwidth}
		\centering
		\includegraphics[width=0.6\textwidth]{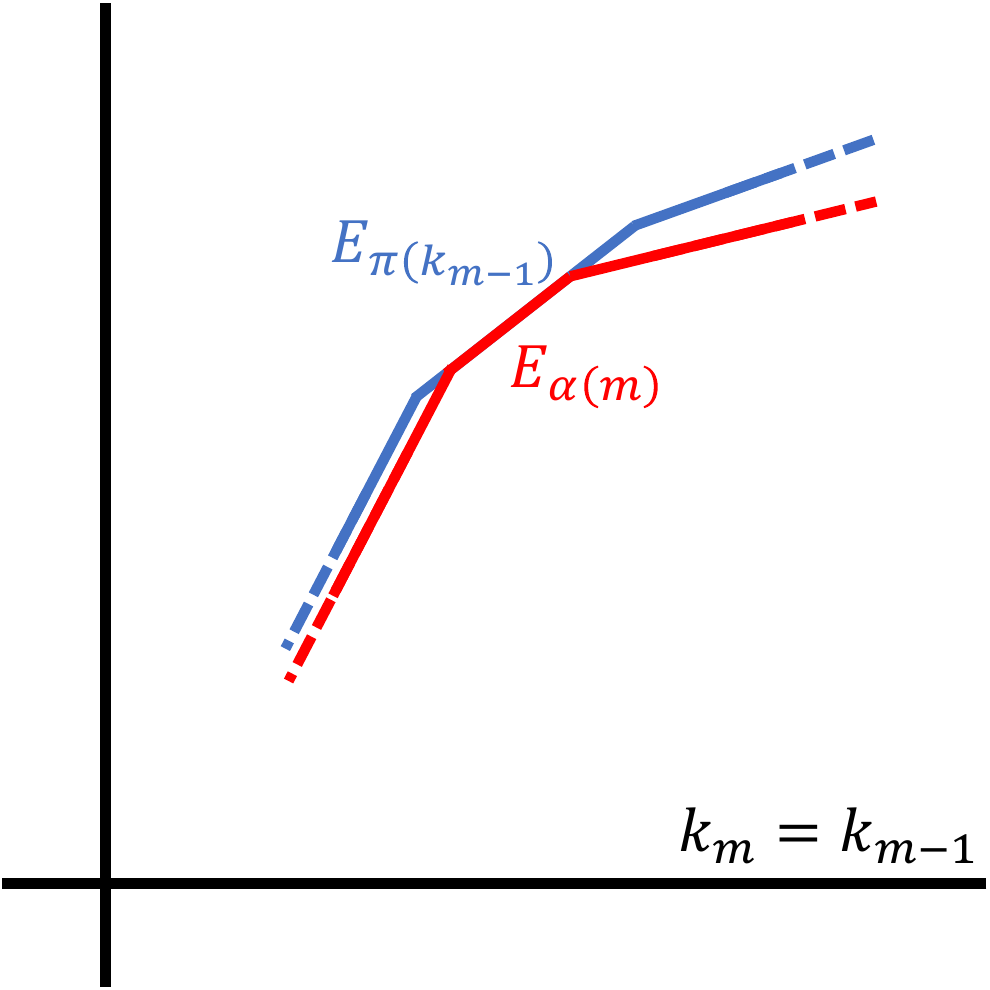}
		\label{fig:rowm case1}
		\centerline{(a) Row $m$, case 1}
	\end{minipage}
	\begin{minipage}{.5\columnwidth}
		\centering
		\includegraphics[width=.6\textwidth]{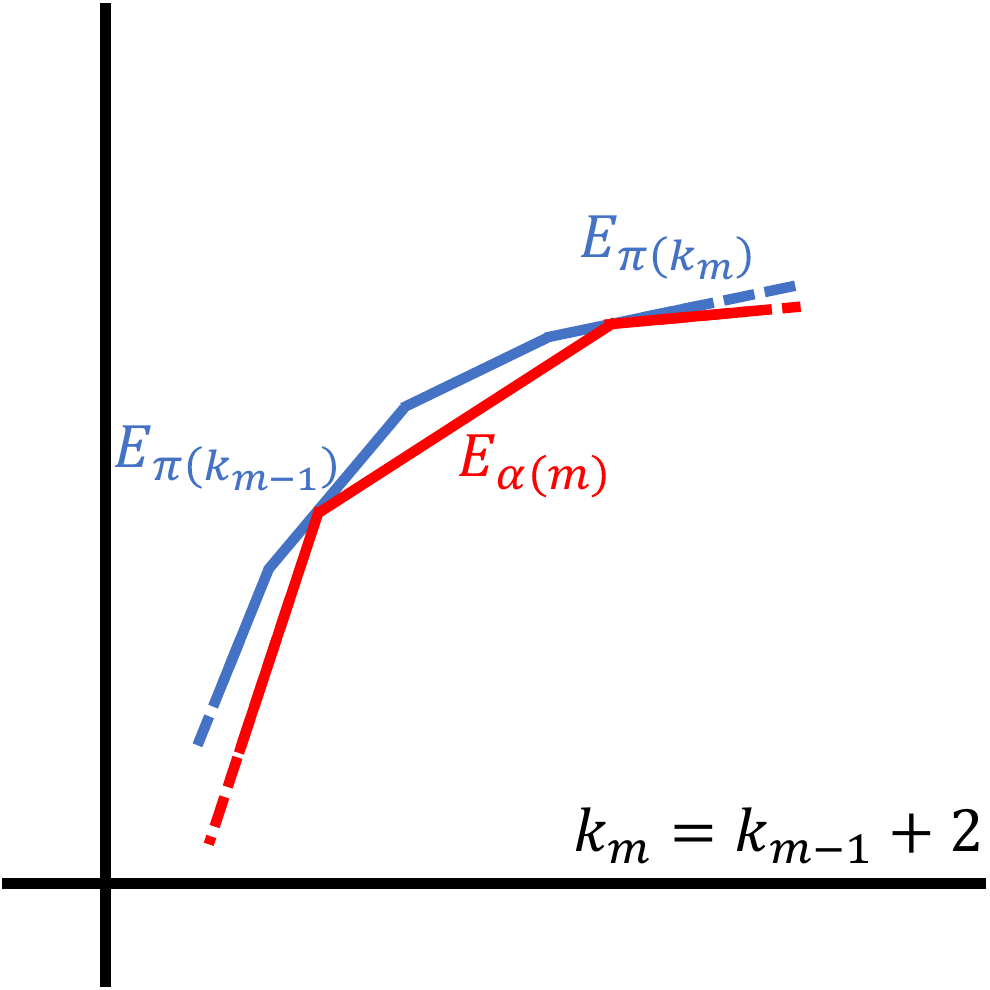}
		\label{fig:rowm case2}
		\centerline{(b) Row $m$, case 2}
	\end{minipage}
	\caption{Thermo-majorization diagrams illustrating the construction of the row $m$ of $P^{(\pi, \alpha)}$. In case 1, \mbox{$\sum_{i=0}^{m}e^{-\beta E_{\alpha\left(i\right)}}<\sum_{j=0}^{k_{m-1}}e^{-\beta E_{\pi\left(j\right)}}$} holds, while in case 2 it does not.}
	\label{fig:rowm}
\end{figure}

In the $m$th row we are determining the population of $E_{\alpha\left(m\right)}$ after the transformation. At this point in the construction, all of the population from energy level $E_{\pi\left(j\right)}$ for $0\leq j \leq k_{m-1}-1$ has been transferred already into the set of energies $\left\{E_{\alpha\left(i\right)}\right\}_{i=0}^{m-1}$. We thus have $G_{mj}=0$ for $0\leq j \leq k_{m-1}-1$. We now check to see how much of the remaining population in $E_{\pi\left(k_{m-1}\right)}$ we can move to $E_{\alpha\left(m\right)}$ subject to the thermo-majorization constraints. If we can only move a fraction of it, we follow the `if' clause in the above algorithm and determine $G_{m k_{m-1}}$ to be given by Eq.~\eqref{eq:G if}. This is illustrated in Fig.~\ref{fig:rowm}a. Alternatively, if we can move all of it, we follow the `else' clause and $G_{m k_{m-1}}$ is given by Eq.~\eqref{eq:m stoc2}. The rest of the construction follows a similar line of argument to the first row. We move all of the population from the energy levels after $E_{\pi\left(k_{m-1}\right)}$ to $E_{\alpha\left(m\right)}$ until we reach an energy level $E_{\pi\left(k_{m}\right)}$, where this is not possible due to thermo-majorization. This is illustrated in Fig.~\ref{fig:rowm}b.

Let \mbox{$c_{\v{p}}:\left[0,Z_S\right]\rightarrow\left[0,1\right]$} be the function such that $c_{\v{p}}(x)$ is the height of the thermo-majorisation curve of $\v{p}$ at $x$. The action of the $\beta$-permutation matrix described above is such that the thermo-majorization curve of $\v{p}^\alpha = P^{(\pi,\alpha)}\v{p}$ is constructed as follows. For $i\in\left\{0,\dots,d-1\right\}$, denoting by $(x^\alpha_i, y^\alpha_i)$ the points that are piecewise linearly connected to give the thermo-majorization curve of $\v{p}^\alpha$, one has:
\begin{enumerate}
	\item Let $x_i^{\alpha}=\sum_{j=0}^{i} e^{-\beta E_{\alpha^{-1}\left(j\right)}}$ and $y_i^{\alpha}=c_{\v{p}}\left(x_i^{\alpha}\right)$.\label{max_beta_condition1}
	\item Define $p^{\alpha}_i:=y^\alpha_{\alpha(i)} - y^{\alpha}_{\alpha(i-1)}$, with $y_{\alpha(-1)}:=0$.\label{max_beta_condition2}
\end{enumerate}
By construction, there is no $\v{q}$ with $\beta$-order $\alpha$ such that $\v{p} \succeq_{th} \v{q} \succeq_{th} \v{p}^\alpha$. 

\subsubsection{Proof of claim 3}

That all extremal points of the thermal polytope associated to $\v{p}$ have the above form is stated and proved in Lemma~12 of Ref.~\cite{lostaglio2018elementary}.

\subsection{Proof of Lemma~\ref{le:active}: optimal unitary control}
\label{ap:lemmaoptimalcontrol}
 First we recall the notion of \emph{maximally active state} associated to a given $\rho$, defined as:
\begin{definition}[Maximally active state]
	Let $\rho$ be the state of a system with associated Hamiltonian \mbox{$H=\sum_{i=0}^{d-1} E_i \ketbra{i}{i}$}, where $E_i\leq E_{i+1}$. The maximally active state associated to $\rho$ is
	\begin{equation*}
	\hat{\rho}=\sum_{i=0}^{d-1}\lambda^\uparrow_i \ketbra{i}{i},
	\end{equation*}
	where $\left\{\lambda^\uparrow_i\right\}_{i=0}^{d-1}$ are the eigenvalues of $\rho$ arranged in ascending order.
\end{definition}
This corresponds to the state with the highest energy along the whole unitary orbit of $\rho$, i.e. $\hat{\rho} = \arg \max_{U} \tr[U \rho U^\dag H]$. We can now prove Lemma \ref{le:active} (note that we use facts from Section \ref{ap:thermalpolytope} in this proof):
\begin{proof}
	The claim is equivalent to proving $\hat{\rho} \succeq_{\textrm{th}} U\rho U^{\dagger}$ for every unitary $U$. We will first show that for any permutation $\pi$ of $\{1,...,d\}$:
	\begin{equation} \label{eq:perm trans}
	\v{\lambda}^{\uparrow} \succeq_{\textrm{th}} \pi \v{\lambda}^{\uparrow},
	\end{equation}
	where $\v{\lambda}^{\uparrow}$ denotes the vector of eigenvalues of $\hat{\rho}$ arranged in ascending order.	To show it, we will construct a sequence of $d-1$ Gibbs-stochastic matrices converting $\v{\lambda}^{\uparrow}$ into $\pi \v{\lambda}^{\uparrow}$, passing through the intermediate states $\v{q}^{\left(1\right)},\dots,\v{q}^{\left(d-1\right)}$, with $\v{q}^{\left(d-1\right)}=\pi \v{\lambda}^{\uparrow}$ and $\v{q}^{\left(0\right)}=\v{\lambda}^{\uparrow}$. 
	
	We start with $\v{q}^{\left(0\right)}$. We wish to move the probability $\lambda^\uparrow_0$ associated with energy level $E_0$ to energy level $E_{\pi\left(0\right)}$. To do this, we transpose in sequence the populations of energy levels \mbox{$0 \leftrightarrow 1$}, \mbox{$1 \leftrightarrow 2$}, $\dots$, \mbox{$\pi(0)-1 \leftrightarrow \pi(0)$}. Each of these is possible under Gibbs-stochastic matrices, because $\lambda^\uparrow_0\leq \lambda^\uparrow_{j+1}$, $\forall j$ and $E_j \leq E_{j+1}$. In fact, they are achieved with the matrices:
	\begin{equation*}
	G_{\left(j,j+1\right)}=\begin{pmatrix}
	1- \mu e^{-\beta\left(E_{j+1}-E_{j}\right)} &  \mu \\
	\mu e^{-\beta\left(E_{j+1}-E_{j}\right)} & 1-\mu
	\end{pmatrix}\oplus\mathbb{I}_{\backslash \left(j,j+1\right)},
	\end{equation*}
	where $\mathbb{I}_{\backslash \left(j,j+1\right)}$ denotes the identity on all levels different from $(j,j+1)$ and
	\begin{equation*}
	\mu=\frac{\lambda^\uparrow_{j+1}-\lambda^\uparrow_0}{\lambda^\uparrow_{j+1}-\lambda^\uparrow_0 e^{-\beta\left(E_{j+1}-E_j\right)}} \geq 0.
	\end{equation*}
	Hence it is always possible to move population $\lambda^\uparrow_0$ to energy level $E_{\pi\left(0\right)}$ by setting
	\begin{equation*}
	\v{q}^{\left(1\right)} = G_{(\pi(0)-1,\pi(0))} \dots G_{(1,2)} G_{(0,1)} \v{q}^{(0)}.
	\end{equation*}
	$\v{q}^{\left(1\right)}$ coincides with $\pi \v{\lambda}^\uparrow$ on element $\pi(0)$.  Now, if we truncate element $\pi(0)$ from $\v{q}^{\left(1\right)}$, we obtain a vector satisfying
	\begin{align*}
	&q^{\left(1\right)}_0=\lambda^\uparrow_1 \leq  q^{\left(1\right)}_1=\lambda^\uparrow_2 \leq\dots \\& \leq q^{\left(1\right)}_{\pi(0)-1} \leq   q^{\left(1\right)}_{\pi(0)+1} \leq\dots \leq  q^{\left(1\right)}_{d-1}=\lambda^\uparrow_{d-1}.
	\end{align*}
	Reasoning as before, we find a second sequence of Gibbs-stochastic matrices acting on the truncated vector, that transposes the populations in adjacent positions and moves $\lambda^\uparrow_1$ to the energy level $E_{\pi\left(1\right)}$. Applied to $\v{q}^{(1)}$, this sequence give a state $\v{q}^{\left(2\right)}$ which coincides with $\pi \v{\lambda}^\uparrow$ in elements $\pi(0), \pi(1)$. It should be clear that this construction can be repeated on every intermediate $\v{q}^{(m)}$, each time applying it to the distribution in which we ignore the energy levels $E_{\pi(0)}, ..., E_{\pi(m-1)}$, since these have the correct occupation probability. This provides a sequence of states culminating in $\v{q}^{\left(d-1\right)}=\pi \v{\lambda}^\uparrow$ as required.
	
	Having shown that Eq.~\eqref{eq:perm trans} holds, we now argue for  $\hat{\rho} \succeq_{\textrm{th}} U\rho U^{\dagger}$ for all unitaries $U$. Given $U$, let $\tilde{\v{p}}=\textrm{diag}\left[U\rho U^\dagger\right]$ (remembering to first diagonalize the degenerate energy subspaces if necessary) and we want to prove $\v{\lambda}^\uparrow \succeq_{th} \tilde{\v{p}}$. Let
	$\tilde{\v{p}}^{\uparrow}$ be the state formed by arranging the elements of $\tilde{\v{p}}$ in ascending order. By Eq.~\eqref{eq:perm trans}, we have that $\tilde{\v{p}}^{\uparrow} \succeq_{\textrm{th}} \tilde{\v{p}}$. In addition, by the Schur-Horn theorem, we have  that $\v{\lambda}^\uparrow \succeq  \tilde{\v{p}}^{\uparrow}$. Combining this with the fact that both $\v{\lambda}^\uparrow$ and $\tilde{\v{p}}^{\uparrow}$ are ordered in terms of increasing occupation probability and have the same $\beta$-order, it follows that:
	\begin{enumerate}
		\item If the thermo-majorization curve of $\v{\lambda}^\uparrow$ is constructed from points $\left\{(x_m, y_m)\right\}_{m=0}^{d-1}$ and that of  $\tilde{\v{p}}^{\uparrow}$ from points $\left\{(x'_m, y'_m)\right\}_{m=0}^{d-1}$, we have $x_m = x'_m$ for each $m$.
		
		\item $y_m \geq y'_m$ for each $m$, since $\sum_{i=0}^{m} \lambda^\uparrow_i \geq \sum_{i=0}^{m} \tilde{p}^\uparrow_i$ for all $m=1,\dots, d-1$.
	\end{enumerate}
	Hence $\v{\lambda}^\uparrow \succeq_{th} \tilde{\v{p}}^\uparrow \succeq_{th} \tilde{\v{p}}$, which implies the claim.
\end{proof}

\subsection{Proof of Theorem~\ref{th:dlevel}}
\label{ap:prooftheorem1}

First, let us focus on optimality in a single round. That the initial unitary can be taken to be the one mapping $\rho^{\left(k-1\right)}_{S}\otimes\rho_A$ to the corresponding maximally active state follows immediately from Lemma~\ref{le:active}. Next, we perform the dephasing thermalization that maximizes the population of the ground state of $S$ within the thermal polytope. If $\pi_k$ is the $\beta$-order of the state after the unitary, this can be chosen to be the $\beta$-permutation $\Lambda^{(\pi_k,\alpha)}$ with $\alpha$ given in Eq.~\eqref{eq:alpha}. That this is an ordering that maximizes the ground state population of $S$ follows from the concavity of thermo-majorization curves. Furthermore, it maximizes $\sum_{i=0}^{l} p^{\downarrow \left(k\right)}_i$, $l\in\left\{0,\dots,d-1\right\}$ where $\left\{p^{\downarrow \left(k\right)}_i\right\}_{i=0}^{d-1}$ are the populations of $\rho^{\left(k\right)}_S$ arranged in descending order. This follows from Lemma~\ref{le:maxbeta} and the fact that the $\beta$-permutation of Eq.~\eqref{eq:alpha} is the one that maximises the $x$-axis coordinates of the elbow points of the output thermo-majorization curve.

We now formally show that the concatenation of such rounds forms an optimal protocol in $\mathcal{P}_{\rho_A}$. Suppose that at the beginning of round $k$ we have one of two diagonal states $\rho^{\left(k-1\right)}_S$ and $\tilde{\rho}^{\left(k-1\right)}_S$ such that $\rho^{\left(k-1\right)}_S\succeq \tilde{\rho}^{\left(k-1\right)}_S$. $\rho^{(k)}_S$ will represent the trajectory followed by the state when we apply the claimed optimal protocol, whereas $\tilde{\rho}^{{\left(k\right)}}_S$ will be the trajectory followed by a generic protocol (hence, $\rho^{(0)}_S = \tilde{\rho}^{(0)}_S$). Our goal is to show that:
\begin{align} \label{eq:maj pres}
\rho^{\left(k\right)}_S&=\tr_A\left[\Lambda^{(\pi_k,\alpha)} \circ\mathcal{U}_{\textrm{m.a.}}^{\left(k\right)}\left(\rho^{\left(k-1\right)}_S\otimes \rho_A\right)\right] \succeq
\tilde{\rho}^{\left(k\right)}_S=\tr_A\left[\Lambda^{\left(k\right)}\circ\mathcal{V}^{\left(k\right)}\left(\tilde{\rho}^{\left(k-1\right)}_S\otimes \rho_A\right)\right]
\end{align}
for all choices of $\rho_A$, $\mathcal{V}^{(k)}$ and $\Lambda^{(k)}$. Here      $\mathcal{U}_{\textrm{m.a.}}^{\left(k\right)}$ denotes the unitary creating the maximally active state on $SA$, $\mathcal{V}^{\left(k\right)}$ is an arbitrary unitary and $\Lambda^{\left(k\right)}$ a dephasing thermalization. This implies not only that our protocol maximizes the ground state population in a given round but also that deviating from it can only have an adverse effect on the achievable population in subsequent rounds.

To see that Eq.~\eqref{eq:maj pres} holds, first note that as \mbox{$\rho^{\left(k-1\right)}_S\succeq \tilde{\rho}^{\left(k-1\right)}_S$}, then:
\begin{equation} \label{eq:joint maj}
\rho^{\left(k-1\right)}_S\otimes \rho_A\succeq \tilde{\rho}^{\left(k-1\right)}_S\otimes \rho_A, \quad \forall \rho_A.
\end{equation}
This gives:
\begin{align} \label{eq:tt}
\mathcal{U}_{\textrm{m.a.}}^{\left(k\right)}\left(\rho^{\left(k-1\right)}_S\otimes \sigma_A\right) \succeq_{\textrm{th}} \mathcal{U}_{\textrm{m.a.}}^{\left(k\right)}\left(\tilde{\rho}^{\left(k-1\right)}_S\otimes \sigma_A\right) \succeq_{\textrm{th}} \mathcal{V}^{\left(k\right)}\left(\tilde{\rho}^{\left(k-1\right)}_S\otimes \sigma_A\right), \quad \forall \mathcal{V}^{\left(k\right)} .
\end{align}
The first line follows from Eq.~\eqref{eq:joint maj} by comparison of the thermomajorization curves, since maximally active states have the same $\beta$-orders. The second line follows from Lemma~\ref{le:active}. Now, by definition of $\Lambda^{(\pi_k,\alpha)}$ and Eq.~\eqref{eq:tt},
\begin{align*}
\tr_A &\left[ \Lambda^{(\pi_k,\alpha)}\circ\mathcal{U}_{\textrm{m.a.}}^{\left(k\right)}\left(\rho^{\left(k-1\right)}_S\otimes \sigma_A\right)\right] \succeq \tr_A \left[ \Lambda^{\left(k\right)}\circ\mathcal{V}^{\left(k\right)}\left(\tilde{\rho}^{\left(k-1\right)}_S\otimes \sigma_A\right)\right].
\end{align*}

\subsection{A qudit protocol with $\beta$-swaps and no ancillas}
\label{ap:protocol}
The $\beta$-permutations affecting only two energy levels $(i,j)$ at once are called \emph{$\beta$-swaps} \cite{lostaglio2018elementary} and have the form
\begin{equation}
\label{eq:betaswapmatrix}
\beta_{i,j}=\begin{pmatrix}
1-e^{-\beta \left(E_j-E_i\right)} & 1\\
e^{-\beta \left(E_j-E_i\right)} & 0
\end{pmatrix}\oplus \mathbb{I}_{\backslash \left(i,j\right)}.
\end{equation}
where $\mathbb{I}_{\backslash (i,j)}$ is the identity on every level $l \neq i,j$. 

We now define the following Gibbs-stochastic matrix, which is a many-level generalization of the matrix $A_6$ from Ref.~\cite{mazurek2017preparation}:
\begin{align*}
\tilde{G}=\begin{pmatrix}
1-e^{\beta \Delta_{01}}  & 1-e^{\beta \Delta_{12}}  & ... & 1-e^{\beta \Delta_{d-1 \, d-2}} & 1 \\ 
e^{\beta \Delta_{01}} & 0 & ... & 0 & 0 \\  
0 & e^{\beta \Delta_{12}} & \ddots & 0 & 0 \\  
\vdots & \ddots & \ddots & \ddots & \vdots \\ 
0 & 0 &  ... & 0 & 0 \\
0 & 0 &  ... & e^{\beta \Delta_{d-1\, d-2}} & 0
\end{pmatrix}.
\end{align*}
where $\Delta_{ij} = E_i - E_j$. $\tilde{G}$ can be generated as a sequence of $\beta$-swaps: $ \tilde{G}=\prod_{i=0}^{d-2}\beta_{i+1,i}$. Now consider the protocol in which, at each round, we perform the following steps:
\begin{enumerate}
	\item A unitary transformation $U$ is performed, that swaps the population of the ground state and the most excited state.
	\item $\tilde{G}$ is performed.
\end{enumerate}
Note that $U$ induces the permutation $S_{(0,d-1)}$ that flips the states \mbox{$0 \leftrightarrow d-1$}. Next, define the resulting cooling stochastic matrix $C =\tilde{G} S_{(0,d-1)}$. Also denote by $\Omega = \sum^{d-2}_{i=0} (E_{i+1} - E_i)=E_{d-1}-E_0$. Then one can verify:
\begin{align*}
C^{d-1}=\begin{pmatrix}
1  & 1-e^{-\beta \Omega} & 1-e^{-\beta \Omega}  & \dots  & 1-e^{-\beta \Omega} &  \\
0 & e^{-\beta \Omega}  & 0 & \dots  &0 \\ 
0 & 0 &  e^{-\beta \Omega} & \ddots  & 0 \\
\vdots & \vdots & 0 & \ddots  & \vdots \\
0 & 0 &  \dots & \dots & e^{-\beta \Omega}
\end{pmatrix}.
\end{align*}
Let $p^{(k)}_i$ be occupations after $k$ steps of the protocol, and $R^{(k)} = 1-p^{(k)}_0$. One has
\begin{equation*}
p^{(k \left(d-1\right))}_0 = 1- R^{(k \left(d-1\right))}, \quad R^{(k \left(d-1\right))} = e^{-\beta \Omega k} S^{(1)}.
\end{equation*}
By assumption  $\Omega >0$ and $\beta>0$, and Eq.~\eqref{eq:ddimensional} follows.

\subsection{Proof of trivial Markovian cooling for $d=2$}
\label{ap:markov}

Markovian dephasing thermalizations are defined as those dephasing thermalizations that can be written as the solution of a master equation with a (possibly time-dependent) generator in Lindblad form. Here we show that the lowest achievable temperature by Markovian dephasing thermalizations without ancillary systems is limited to that of the environment.

As discussed in Sec.~\ref{ap:thermalpolytope} of the Methods, dephasing thermalizations act on the population vector $\v{p}$ as Gibbs-stochastic matrices, in this case $2 \times 2$. Using the results of Ref.~\cite{wolf2008assessing}, a direct computation shows that the action on $\v{p}$ of Markovian dephasing thermalizations can be written as
\begin{equation}
G_{\rm Mark} = (1-\lambda)\mathbb{I} +  \lambda \beta_{01}, \quad 0 \le \lambda \le 1/(1+e^{-\beta E})
\end{equation}
with $\beta_{01}$ given by Eq.~\eqref{eq:betaswapmatrix}. 

Let $(p,1-p)$, $(q,1-q)$ $(s,1-s)$ be the energy distribution at the start, after the unitary, and after the Markovian dephasing thermalization, respectively. A direct calculation shows that the thermalization relates $\v{s}$ and $\v{q}$ by
\begin{equation}
s=\left(1-\lambda e^{-\beta E}\right)q+\lambda (1-q).
\end{equation}
Since $\v{p}$ and $\v{q}$ are related by a unitary, we have that
\mbox{$p(1-p) \le q(1-q)$}. Assume that the system is initially hotter than the bath, $p < 1/(1+e^{-\beta E})$. By direct inspection the maximum is found at $\lambda=1/(1+e^{-\beta E})$, where one achieves $s=1/(1+e^{-\beta E})$, the thermal ground state population at temperature $\beta$. The same would be true if one optimized over all Markovian thermal operations (using again the results of Ref.~\cite{wolf2008assessing} and the proof in Sec.~\ref{ap:opt qubit}).

\subsection{Proof of Theorem~\ref{th:reusability}}
\label{ap:theorem2}

\begin{figure}[t!]
	\centering
	\includegraphics[width=0.8\linewidth]{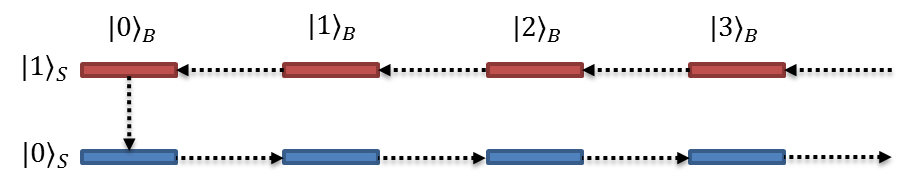}
	\caption{The circulation of populations induced by each round (Pauli $X$ followed by $U^{\beta}_{SB}$) of the optimal protocol. Arrows indicate complete transfer of population. The picture gives an intuitive understanding of the cooling mechanism of the optimal protocol.} 
	\label{fig:circulation}
\end{figure}

We prove this theorem via a direct calculation of the ground state probability after a repeated application of the unitaries.
Note that the state $\rho^{(k)}_S$ at the beginning of round $k+1$ of the optimal protocol is incoherent for every $k \geq 1$. As such, the system-bath state after round $k$ will have the general form:
\begin{equation*}
\rho^{\left(k\right)}_{SB}=\sum_{n=0}^{\infty} \left(p^{\left(k\right)}_{0,n} \ketbra{0,n}{0,n}+p^{\left(k\right)}_{1,n}\ketbra{1,n}{1,n}\right),
\end{equation*}
for some occupation probabilities $p^{(k)}_{i,n}$ with $i\in\left\{0,1\right\}$ and $n \in \left\{0,1,\dots\right\}$. 
Applying $X$ to $S$, followed by  $U^{\beta}_{SB}$ gives
\begin{small}
	\begin{align*}
	&\rho^{\left(k\right)}_{SB}\mapsto  \rho^{\left(k+1\right)}_{SB}=p^{\left(k\right)}_{1,0}\ketbra{0,0}{0,0}+p^{\left(k\right)}_{0,0}\ketbra{0,1}{0,1}
	\\ &
	+\sum_{n=1}^{\infty}\left(p^{\left(k\right)}_{1,n}\ketbra{1,n-1}{1,n-1}+p^{\left(k\right)}_{0,n}\ketbra{0,n+1}{0,n+1}\right).
	\end{align*}
\end{small}
The relation between the occupations at step $k$ and those at step $k+1$ is (see Fig.~\ref{fig:circulation})
\begin{align*}
p^{\left(k+1\right)}_{0,0}&=p^{\left(k\right)}_{1,0},\\
p^{\left(k+1\right)}_{0,n}&=p^{\left(k\right)}_{0,n-1},\quad\quad n\geq 1,\\
p^{\left(k+1\right)}_{1,n}&=p^{\left(k\right)}_{1,n+1},\quad\quad n\geq0.
\end{align*}

Solving these equations recursively, we find the populations at step $k$ as a function of the initial occupation probabilities (at $k=0$):
\begin{align*}
p^{\left(k\right)}_{1,n}&=p^{\left(0\right)}_{1,n+k}\\
p^{\left(k\right)}_{0,n}&=\begin{cases}
p^{\left(0\right)}_{0,n-k}, \quad & n\geq k,\\
p^{\left(0\right)}_{1,k-n+1}, \quad & n<k.
\end{cases}
\end{align*}
As before, we denote by $p^{(k)}_0$ the ground state population of the system after $k$ steps of the protocol. We have
\begin{small}
	\begin{align*}
	p^{(k)}_0 = p_{0,0}^{(k)} + \sum_{n=1}^{\infty} p^{(k)}_{0,n}= p^{(k-1)}_{1,0} + \sum_{n=1}^{\infty} p^{(k-1)}_{0,n-1}= p^{(k-1)}_{1,0} + p^{(k-1)}_0.
	\end{align*}
\end{small}
We now use the relation
\begin{equation*}
p^{(k-1)}_{1,0} = p^{(k-2)}_{1,1} = \dots = p^{(0)}_{1,k-1} = p^{(0)} t^{(0)}_k,
\end{equation*}
where $t^{(0)}_k=\left(1-e^{-\beta E}\right)e^{-k\beta E}$ denotes the occupation of the $k$th energy level of the bosonic mode at the beginning of the protocol. This finally leads to
\begin{equation} \label{eq:no refresh}
p^{\left(k\right)}_0=p^{\left(0\right)}_0+\left(1-p^{\left(0\right)}\right)\sum_{n=0}^{k-1} t^{\left(0\right)}_{n}.
\end{equation}
Direct substitution shows that this expression is identical to that derived in Theorem~\ref{th:qubitprot} for the optimal protocol.

\subsection{Robustness of Theorem \ref{th:reusability} to anharmonicity}\label{ap:anharmonic}
\label{ap:anharmonicity}
We model an anharmonicity in the oscillator as a correction to the energy gap between the energy levels $n$ and $n+1$ (which we now label as $E^{\text{an}}_{n}$)
\begin{equation}\label{eq:anhcorr}
E^{\text{an}}_{n+1}- E^{\text{an}}_n = E(1-(n+1)\tau^2) + o(\tau^2),
\end{equation}
where $\tau \ll 1$.
To analize its effect, notice that the ground state from Eq. \eqref{eq:no refresh} is
\begin{equation} \label{eq:no refresh2}
p^{\left(k\right)}_0=p^{\left(0\right)}_0+\left(1-p^{\left(0\right)}\right)\sum_{n=0}^{k-1} t^{\left(0\right)}_{n},
\end{equation}
and that this does not rely on any particular form for the populations $t_k^{(0)}$. Thus, for any $\tau$, we have to compare the two following sums,
\begin{equation}
\sum_{n=0}^{k-1} t^{\left(0\right),\text{an}}_{n} =\frac{\sum_{n=0}^{k-1}e^{-\beta E^{\text{an}}_n}}{\sum_{n=0}^{\infty}e^{-\beta E^{\text{an}}_n}} \quad 
\sum_{n=0}^{k-1} t^{\left(0\right)}_{n} =\frac{\sum_{n=0}^{k-1}e^{-\beta E_n}}{\sum_{n=0}^{\infty}e^{-\beta E_n}}.
\end{equation}

A significantly large value $\tau = 0.05$ gives a distortion $E^{\rm an}_n/E_n$ of $5 \% $ at $n=8$. Also, $\frac{\sum_{n=0}^{k-1} t^{\left(0\right),\text{an}}_{n}}{\sum_{n=0}^{k-1} t^{\left(0\right)}_{n}}$ approaches $1$ as $k$ grows and the largest deviation from $\frac{\sum_{n=0}^{k-1}e^{-\beta E_n}}{\sum_{n=0}^{\infty}e^{-\beta E_n}} $ peaks at less than $0.005 \%$. This shows that these protocols are robust to realistic anharmonicities.

\subsection{Proof of Theorem \ref{thm:robustness} and its extension to thermal operations} \label{ap:opt qubit}

We give a proof of a more general result than Theorem \ref{thm:robustness}, optimizing the thermalization step over the set of thermal operation (TO) $\mathcal{T}$ on a qubit, which strictly include all dephasing thermalizations.

\subsubsection{Preliminaries}\label{ap:prelim}

First we review the definition of thermal operations \cite{janzing2000thermodynamic, brandao2011resource}, which are channels acting on a quantum system $S$ in state $\rho_S$ and with Hamiltonian $H_S$ as:
\begin{equation} \label{eq:TO def}
\mathcal{T} \left(\rho_S\right)=\tr_B\left[U\left(\rho_S\otimes\frac{e^{-\beta H_B}}{\tr\left[e^{-\beta H_B}\right]}\right)U^\dagger\right],
\end{equation}  
where $\beta= 1/(kT)$ is the inverse temperature of the surrounding heat-bath, $H_B$ is an arbitrary Hamiltonian and $U$ is an energy preserving unitary that satisfies $\left[U,H_S+H_B\right]=0$. 

Let $S$ be a qubit with Hamiltonian $H_S=E\ketbra{1}{1}$. Given $\rho_S$, in addition to the occupation probabilities $\v{p}_S=\left(p,1-p\right)^T$, let $\rho_{01}=\bra{0}\rho_S\ket{1}$ denote the off-diagonal term. Similarly, for a second state $\sigma_S$, let $\v{q}_S = \left(\bra{0}\sigma_S\ket{0}, \bra{1}\sigma_S\ket{1}\right)^T$ and $\sigma_{01}=\bra{0}\sigma_S\ket{1}$. Then any thermal operation on a qubit can be described by:
\begin{enumerate}
	\item The action on the population (given by a Gibbs-stochastic matrix $G$).
	\item The action on the off-diagonal term.
\end{enumerate}
Taken together this gives:
\begin{equation}
\label{eq:thermaloperationparamters}
\mathcal{T}(\rho_S) = \sigma_S \, \Leftrightarrow
\begin{cases}
\v{q_S} = G \v{p}_S =  \begin{pmatrix}
1-\lambda e^{-\beta E} & \lambda \\
e^{-\beta E} \lambda & 1-\lambda \end{pmatrix}\v{p}_S, \quad \lambda \in [0,1]\\ 
|\sigma_{01}| = c |\rho_{01}|, \quad  0 \leq c\leq  \sqrt{(1-e^{-\beta E}\lambda)(1-\lambda)}
\end{cases},
\end{equation}
where the expression for $c$ follows from the constraint $|c| \leq \sqrt{G_{00}G_{11}}$ \cite{cwiklinski2015limitations, lostaglio2015quantum} which is a consequence of the complete positivity of $\mathcal{T}$. 
One can also apply a phase $e^{i \theta}$ to $\rho_{01}$ by a unitary, so that in general $c \not \in \mathbb{R}$, but since this will be irrelevant in the following considerations, without loss of generality we set $\theta =0$ and take $c\geq 0$ (one can always reversibly transform to any $\theta \neq 0$ by an energy-preserving unitary, which is a thermal operation). Hence, for our purposes a qubit thermal operation $\mathcal{T}$ is defined by the two parameters:
\begin{equation}
\label{eq:thermalopparameterization}
\v{v}\left[\mathcal{T}\right] = \left(\lambda, c\right).
\end{equation} 
With this notation, the $\beta$\emph{-swap} introduced in the main text  is a thermal operation $\mathcal{T}_{\beta}$ that removes all quantum coherences and has $G=G^{\beta\textrm{-swap}}$ where
\begin{equation} \label{eq:betaswap}
G^{\beta\textrm{-swap}}=
\begin{pmatrix}
1-e^{-\beta E} & 1\\
e^{-\beta E} & 0
\end{pmatrix}
\end{equation}
corresponds to $\v{v}\left[\mathcal{T}_{\beta}\right] = \left(1, 0\right)$. Dephasing thermalizations are the subset of thermal operations with $c=0$.

\subsubsection{Proof of an extended version of Theorem~3} 
Define the set of $\epsilon$-noisy thermal operations as the set of thermal Operation $\mathcal{T}_\epsilon$ such that 
\begin{equation}
\v{v}[\mathcal{T}_\epsilon] = (\lambda,c), \quad \lambda \leq 1-\epsilon.
\end{equation}
Let $\mathcal{T}^\epsilon_\emptyset$ denote the set of cooling protocols using no ancilla in which at each round $k$
\begin{enumerate}
	\item A unitary $U^{(k)}$ is applied to $S$,
	\item An $\epsilon$-noisy thermal operation $\mathcal{T}^{(k)}_\epsilon$ is applied to $S$.
	\item A unitary $V^{(k)}$ is applied to $S$ 
\end{enumerate} 
Then, 
\begin{restatable}{theorem}{thmrobustness}
	Under the assumptions of Corollary~\ref{th:qubitprot} and given $\epsilon \leq \frac{1}{1+e^{\beta E} + e^{2\beta E}} $, the optimal nontrivial cooling protocol in $\mathcal{T}^\epsilon_\emptyset$ is such that in each round $k$:
	\begin{enumerate}
		\item The Pauli $X$ unitary is applied to $S$.
		
		\item Any $\epsilon$-noisy thermal operation $\mathcal{T}_\epsilon$ with $\v{v}(\mathcal{T}_\epsilon) = (1-\epsilon,c)$ is applied to $S$.
	\end{enumerate}
	The population of the ground state after round $k$ is:
	\begin{align}
	\label{eq:coolingrobustness}
	p_{\rm 0}^{(k)} =1-\frac{\epsilon}{2-(1-\epsilon) Z}\nonumber 
	\,\,- \left((1-\epsilon) Z -1\right)^k \left( 1-\frac{\epsilon}{2-(1-\epsilon) Z}- p_0^{(0)}\right),
	\end{align}
	where $Z = 1 + e^{-\beta E}$ and $p^{\left( k\right)}_0\rightarrow 1-\frac{\epsilon}{2-(1-\epsilon)Z}$ as $k\rightarrow\infty$.
\end{restatable}
Before we prove this theorem, let us discuss the consequences. Note that a particular choice for the optimal cooling protocol is $\mathcal{T}_\epsilon$ with $\v{v}(\mathcal{T}_\epsilon) = (1-\epsilon,0)$. This is just the $\epsilon$-noisy $\beta$-swap defined in the main text. Since $\mathcal{T}^\epsilon_{\emptyset} \subset \mathcal{P}^\epsilon_{\emptyset}$, this means that the above theorem implies Theorem~3 of the main text as an immediate corollary. Also note, that setting $\epsilon = 0$, we obtain as a simple consequence a direct proof of Corollary~1 that does not goes through Theorem~1. 

Furthermore, note it does not matter what is the choice for $c$ in the $\epsilon$-noisy thermal operation: it could even vary from round to round.  The Jaynes-Cummings implementation described in the main text is an $\epsilon$-noisy thermal operation. On the population it acts as an $\epsilon$-noisy $\beta$-swap, but it has $c \neq 0$. However, the above theorem shows that this does not make any difference, since having control on the coherent part of the evolution does not provide any advantage and the performance is independent by the choice of $c$ at each step. In other words, as claimed in the main text, we can safely ignore the coherent evolution and focus on the induced population dynamics, at least for $\epsilon$ small enough.

\begin{proof}
	The most general protocol can be written as 
	\begin{equation}
	\label{eq:generalprotocol}
	\rho^{(k)}_S = \mathcal{V}^{(k)}\mathcal{T}^{(k)} \mathcal{U}^{(k)} \dots \mathcal{V}^{(1)}\mathcal{T}^{(1)} \mathcal{U}^{(1)}\mathcal{U}_{\textrm{di}}(\rho_S),  
	\end{equation}
	where $\mathcal{U}^{(i)}(\cdot) = U^{(i)}(\cdot)U^{(i)\, \dagger}$ and $\mathcal{V}^{(i)}(\cdot) = V^{(i)}(\cdot)V^{(i)\, \dagger}$ denote general unitaries applied to $S$ in round $i$ and $\mathcal{T}^{(i)}$ is a thermal operation on $S$ applied at round $i$. Here $\mathcal{U}_{\textrm{di}} = U_{\textrm{di}} (\cdot)U^\dag_{\textrm{di}}$ is the unitary that transforms $\rho_S$ into a diagonal form with $p_0^{(0)} \geq p_1^{(0)}$. This can be done without loss of generality, since any protocol in $\mathcal{P}$ starts with an arbitrary unitary $\mathcal{U}^{(1)}$. 
	
	We want to maximize the ground state population over all choices of $\mathcal{U}^{(i)}$, $\mathcal{T}^{(i)}$ and $\mathcal{V}^{(i)}$.  As we perform an arbitrary unitary at the beginning of every round, we can assume without loss of generality that the state of $S$ at the start of round $k+1$ is diagonal in the energy eigenbasis:
	\begin{equation*}
	\rho^{\left(k\right)}_S=\begin{pmatrix}
	p^{\left(k\right)} & 0\\
	0 & 1-p^{\left(k\right)}
	\end{pmatrix}
	\end{equation*}
	and that $p^{\left(k\right)}\geq\frac{1}{2}$ (here we drop the subscript 0 to simplify the notation). In round $k+1$ of the protocol, we have:
	\begin{align*}
	\rho^{\left(k\right)}_S=\begin{pmatrix}
	p^{\left(k\right)} & 0\\
	0 & 1-p^{\left(k\right)}
	\end{pmatrix}
	\stackrel{\mathcal{U}^{\left(k\right)}}{\longrightarrow}&\rho'^{\left(k\right)}_S=\begin{pmatrix}
	q^{\left(k\right)} & a^{(k)}\\
	a^{*(k)} & 1-q^{\left(k\right)}
	\end{pmatrix}\\
	\stackrel{\mathcal{T}^{\left(k\right)}}{\longrightarrow}&\rho''^{\left(k\right)}_S=\begin{pmatrix}
	s^{\left(k\right)} & b^{(k)}\\
	b^{*(k)} & 1-s^{\left(k\right)}
	\end{pmatrix}\\
	\stackrel{\mathcal{V}^{\left(k\right)}}{\longrightarrow}&\rho^{\left(k+1\right)}_S=\begin{pmatrix}
	p^{\left(k+1\right)} & 0\\
	0 & 1-p^{\left(k+1\right)}
	\end{pmatrix}
	\end{align*}
	and our goal is to maximize $p^{\left(k+1\right)}$.
	
	As $\rho^{\left(k+1\right)}_S$ and $\rho''^{\left(k\right)}_S$ are related by a unitary, maximizing $p^{\left(k+1\right)}$ corresponds to minimizing the determinant of $\rho''^{\left(k\right)}_S$. Among all thermal operations associated to a fixed Gibbs-stochastic matrix $G$ with $G\v{q}^{(k)}:=\v{s}^{(k)}$ (with $\v{q}^{\left(k\right)}$ and $\v{s}^{\left(k\right)}$ defined through the above matrices), optimality of the protocol imposes that we choose $\mathcal{T}^{(k)}$ to be a thermal operation that maximizes the absolute value of $b^{(k)}$ - i.e. it preserves the maximum possible amount of coherence. This is achieved by the thermal operation that maximizes the parameter $c$ for given $\lambda$, i.e., from Eq.~\eqref{eq:thermaloperationparamters}, $c= (1- \lambda e^{-\beta E})(1-\lambda)$ or, with the parametrization of Eq.~\eqref{eq:thermalopparameterization}, $\v{v}[{\mathcal{T}^{(k)}}] = (\lambda, (1- \lambda e^{-\beta E})(1-\lambda))$.
	
	What remains to be done is to perform an optimization over all possible Gibbs-stochastic matrices $G$, parametrized by $\lambda \in [0,\lambda_{\max}]$ as in Eq.~\eqref{eq:thermaloperationparamters}, where $\lambda_{\max}\ge 1-\frac{1}{1+e^{\beta E}+e^{2 \beta E}}$ (which corresponds to $\epsilon \le \frac{1}{1+e^{\beta E}+e^{2 \beta E}}$). For each $G$, the relation between $s^{\left(k\right)}$, $b^{(k)}$ and $q^{\left(k\right)}$ and $a^{(k)}$ is
	\begin{align*}
	s^{\left(k\right)} &= \left(1-\lambda e^{-\beta E}\right)q^{\left(k\right)}+\lambda\left(1-q^{\left(k\right)}\right)\\
	|b^{\left(k\right)}  |^2&=|a^{\left(k\right)} |^2\left(1-\lambda e^{-\beta E}\right)\left(1-\lambda\right).
	\end{align*}
	Finally, using the unitarity of $U^{\left(k\right)}$, we can relate $q^{\left(k\right)}$ and $a^{(k)}$ to $p^{\left(k\right)}$ via:
	\begin{equation*}
	p^{\left(k\right)}\left(1-p^{\left(k\right)}\right)=q^{\left(k\right)}\left(1-q^{\left(k\right)}\right)-\left|a^{(k)}\right|^2.
	\end{equation*}
	The determinant of $\rho''^{\left(k\right)}_S$ is thus given by
	\begin{align}
	\label{eq:detfunction}
	f_{p^{\left(k\right)},E}\left(q^{\left(k\right)},\lambda\right)=& \left(\lambda -1\right) \left(e^{-\beta E} \lambda -1\right) \left(-\left(p^{\left(k\right)}-q^{\left(k\right)}\right)\right) \left(p^{\left(k\right)}+q^{\left(k\right)}-1\right)\nonumber\\
	&\quad	-\left(q^{\left(k\right)} \left(e^{-\beta E} \lambda +\lambda -1\right)-\lambda \right)  \left(q^{\left(k\right)} \left(e^{-\beta E} \lambda +\lambda -1\right)-\lambda +1\right),
	\end{align}
	for fixed $E$ and $p^{\left(k\right)}$. Note that the equation is quadratic in both $\lambda$ and $q^{\left(k\right)}$. The coefficient of $\lambda^2$ is
	\begin{equation*}
	-\left[\left(1-\left(q^{\left(k\right)}\right)^2\right)-q^{\left(k\right)} e^{-\beta E}\right]^2 -e^{-\beta E}\left[q^{\left(k\right)}\left(1-q^{\left(k\right)}\right)-p^{\left(k\right)}\left(1-p^{\left(k\right)}\right)\right],
	\end{equation*}
	which is clearly negative as $q^{\left(k\right)}\left(1-q^{\left(k\right)}\right)=p^{\left(k\right)}\left(1-p^{\left(k\right)}\right)+|a^{\left(k\right)} |^2$, and  thus the minimum values will be obtained at either $\lambda = 0$ or $\lambda =\lambda_{\max}$. The case $\lambda=0$ corresponds to not implementing the TO and leads to $p^{\left(k+1\right)}=p^{\left(k\right)}$.
	Taking $\lambda=\lambda_{{\rm max}}$, we still need to show that the Pauli $X$ unitary at each step is optimal.
	
	Let us rewrite the function $f_{p^{\left(k\right)},E}\left(q^{\left(k\right)},\lambda_{\rm max}\right)$ as a polynomial in $q^{(k)}$,
	\begin{equation*}
	f_{p^{\left(k\right)},E}\left(q^{\left(k\right)},\lambda_{\rm max}\right)\equiv f^{(2)}_{p^{\left(k\right)},E}\left(\lambda_{\rm max}\right)(q^{(k)})^2 +f^{(1)}_{p^{\left(k\right)},E}\left(\lambda_{\rm max}\right)q^{(k)}+f^{(0)}_{p^{\left(k\right)},E}\left(\lambda_{\rm max}\right).
	\end{equation*}
	Given that it is a quadratic equation, the location of its minimum depends on the sign of the coefficient $f^{(2)}_{p^{\left(k\right)},E}\left(\lambda_{\rm max}\right)=\lambda_{\rm max}(1+e^{-\beta E}-\lambda_{\rm max}(1+e^{-\beta E}+e^{-2 \beta E}))$. To see when the Pauli $X$ is optimal, we want to find the cases in which the solution is either $q^{(k)}=p^{(k)}$ or $q^{(k)}=1-p^{(k)}$ (that is, on the boundary of the range, given unitarity), which occurs when $f^{(2)}_{p^{\left(k\right)},E}\left(\lambda_{\rm max}\right)< 0$. This is equivalent to
	\begin{equation}\label{eq:maxlambda}
	\lambda_{{\rm max}} > \frac{1+e^{-\beta E}}{1+ e^{-\beta E} + e^{-2\beta E}}=1-\frac{1}{1+e^{\beta E}+e^{2 \beta E}},
	\end{equation}
	which is true by assumption. On top of this, we find that 
	\begin{equation*}
	f_{p^{\left(k\right)},E}\left(p^{\left(k\right)},\lambda_{\rm max}\right)-f_{p^{\left(k\right)},E}\left(1-p^{\left(k\right)},\lambda_{\rm max}\right)= \lambda_{{\rm max}}(1-e^{-\beta E})(\lambda_{{\rm max}} (1+e^{-\beta E})-1)(2 p^{\left(k\right)} -1 ),
	\end{equation*}
	so the minimum is at $q^{(k)}=\min\left\{p^{\left(k\right)},1-p^{\left(k\right)}\right\} = 1-p^{(k)}$ as long as $\lambda_{{\rm max}} > \frac{1}{1+ e^{-\beta E} }$ (which is implied by Eq. \eqref{eq:maxlambda}).
	
	Thus, the optimal protocol in $\mathcal{T}^\epsilon_{\emptyset}$ is one in which, for every $k$, $\mathcal{T}^{(k)}$ is a thermal operation $(\lambda_{{\rm max}}, c_k)$ with $\lambda_\text{max}=1-\epsilon$ (the ``best approximation'' of the $\beta$-swap) and $\mathcal{U}^{(k)}(\cdot) = \mathcal{X}(\cdot) := X(\cdot)X^\dagger$. Note that we do not specify how each  $\mathcal{T}^{(k)}$ acts on the off-diagonal element of the quantum state (i.e., the parameter $c_k$) simply because the input state contains no coherence. As such, the protocol is also optimal for the set of dephasing thermalizations, since at step $k$ one can perform any TO with $G_{01} = \lambda_{{\rm max}}$, without any control required on $c_k$. The optimal ground state population achieved by the above protocol satisfies 
	\begin{equation*}
	p^{\left(k+1\right)}=\left(1-\lambda_{{\rm max}} e^{-\beta E}\right)\left(1-p^{\left(k\right)}\right)+\lambda_{{\rm max}} p^{\left(k\right)},
	\end{equation*}  
	as one can verify by a direct computation.
	Solving this recursion relation gives Eq.~\eqref{eq:coolingrobustness}. One recovers the scaling of Theorem~\ref{th:qubitprot} when $\lambda_{{\rm max}} =1$. Furthermore, since $\lambda_{{\rm max}} Z -1 \leq e^{-\beta E}$ one has exponential convergence to $1-\frac{1-\lambda_{\textrm{max}}}{1-2\lambda_{\textrm{max}}Z}$. Note that the trivial protocol in which we do not do a thermal operation is optimal only when \mbox{$p^{(0)} \geq 1-\frac{1-\lambda_{\textrm{max}}}{1-2\lambda_{\textrm{max}}Z}$} (that is, when the initial ground state population is higher than the optimal asymptotic value).
	
	Finally, let us show that the concatenation of optimal rounds is optimal overall. To this end, let $p^{(k)}$ and $\tilde{p}^{(k)}$ be two ground state populations with $p^{(k)} \geq \tilde{p}^{(k)}$. One can compute
	\begin{equation*}
	p^{(k+1)} - \tilde{p}^{(k+1)} \geq (\lambda_{\textrm{max}}Z -1)(p^{(k)} - \tilde{p}^{(k)}) \geq 0,
	\end{equation*}
	since $\lambda_{\textrm{max}} > 1/Z$. Hence the optimal protocol is a concatenation of the optimal single round protocol, and is also independent of the initial state of the system (with exclusion of the $k=0$ unitary).  
\end{proof}

\section{Experimental proposal}

\subsection{Bounds on the achievable cooling in the Jaynes-Cummings model}
\label{ap:JC}

We compare the performance of a protocol that uses a Jaynes-Cummings (JC) interaction to implement the $\beta$-swap against the ideal unitary $U^{\beta}_{SB}$ of Eq.~\eqref{eq:uSB}. Consider a resonant JC Hamiltonian in rotating wave approximation
\begin{equation}
\label{eq:JCHamiltonian}
H_{\textrm{JC}} = g( \sigma_+ \otimes a + \sigma_- \otimes a^\dag),
\end{equation}
coupling $S$ with a single-mode bosonic bath prepared in a thermal state. Here $a^\dag$ and $a$ are creation and annihilation operators on $B$ and $\sigma_{+} = \ketbra{1}{0}$ and $\sigma_- = \ketbra{0}{1}$. The de-excitation probability is
\begin{equation}
\label{eq:JClambda}
G_{0|1}(s) = \frac{1}{Z_B}\sum_{n=1}^{\infty} \sin^2(s \sqrt{n}) e^{-\beta E (n-1)},
\end{equation}
where $Z_B = (1-e^{-\beta E})^{-1}$ and $s$ is the normalized interaction time ($s = g t$, if $U_{JC}= e^{-i H_{JC}t}$). To realize a dephasing thermalization, strictly speaking one should dephase in the energy basis. However, since in the JC interaction the evolutions of populations and coherences are decoupled and the subsequent Pauli $X$ simply inverts the populations, we can skip this step (furthermore, the results of Sec.~\ref{ap:opt qubit} show that these coherences cannot be exploited by changing the unitary step).  Hence at each round $k$:
\begin{enumerate}
	\item A Pauli $X$ operation is performed on $S$.
	\item The interaction Hamiltonian $H_{\rm JC}$ operates for a time $s$ that maximizes Eq.~\eqref{eq:JClambda}.
	\item The bosonic mode is reset to a thermal state.
\end{enumerate}

In Ref.~\cite{lostaglio2018elementary}, temperature-dependent upper bounds for the maximum achievable transition probability $G^{{\rm max}}(s)$ were derived as a function of $ \bar{\beta} = \beta E$:
\begin{align*}\label{eq:upperbound lambda}
G^{{\rm max}}(\bar{\beta}) \leq\left\{
\begin{array}{ll}
\frac{1}{16} \left(8 e^{-\bar{\beta}}-e^{2 \bar{\beta}}+e^{3 \bar{\beta}}+8\right), \; \textrm{for } \bar{\beta} \in [0, \frac{\log(4)}{3}], \\
e^{-4 \bar{\beta}}-e^{-3 \bar{\beta}}+1, \quad \quad \quad \quad \quad \textrm{for } \bar{\beta} \geq \frac{\log(4)}{3}.
\end{array}
\right.
\end{align*}
Substituting these upper bounds in Theorem ~\ref{thm:robustness}, we obtain upper bounds on the ground state population achieved in the JC models after $k$ rounds, presented in Fig.~\ref{fig:comparison}. In particular for the asymptotic population
\begin{equation}\label{eq:upperbound p_max}
p^{(\infty)}_0 \leq \left\{
\begin{array}{ll}
\frac{1}{e^{\bar{\beta} }+\frac{16 e^{2 \bar{\beta} }}{-16 e^{\bar{\beta} }+e^{3 \beta }-8}+1}, \; \textrm{for } \bar{\beta} \in [0, \frac{\log(4)}{3}], \\
\frac{1}{\frac{e^{\bar{\beta} }}{e^{4 \bar{\beta} }+1}+1}, \quad \quad \quad \quad \quad \textrm{for } \bar{\beta} \geq \frac{\log(4)}{3}.
\end{array}
\right.
\end{equation}
To obtain an explicit protocol whose performance lower bounds the optimal, one can numerically optimize $G_{0|1}(s)$ over $s$ within a finite domain. We take \mbox{$s \in [0,5 \times 10^3]$}, and obtain the curves in Fig.~\ref{fig:comparison}. As an illustrative example, for $\bar{\beta} =1$ one has that the optimal JC asymptotic cooling is $p^{(\infty)}_0 \in [0.9401,0.9534]$. As we know from Theorem~\ref{thm:robustness}, the convergence is exponential and can be computed explicitly from Eq.~\eqref{eq:coolingrobustness} with $\epsilon = 1-G_{0|1}(\tilde{s})$ for the chosen interaction time $\tilde{s}$. 

\begin{figure}[t!]
	\centering
	\includegraphics[width=	0.7\linewidth]{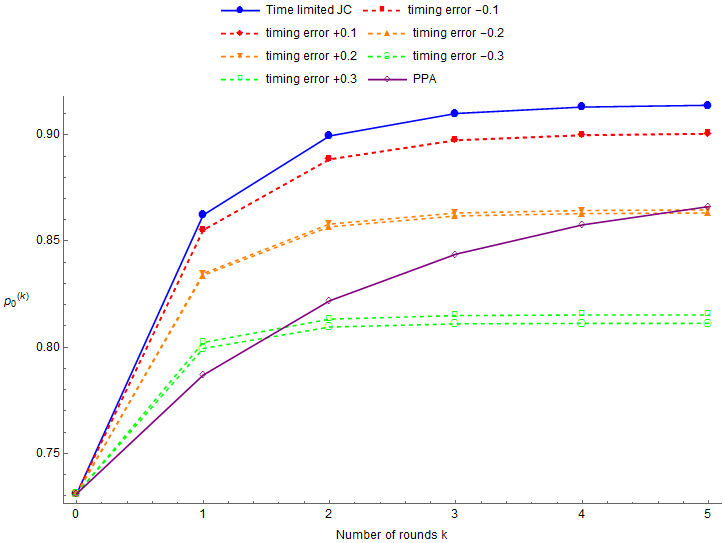}
	\caption{In the time-limited JC model (blue curve on top) we optimise the interaction time $s$ over the limited period $s \in [0,10]$. To this we add increasing errors in the accuracy with which $s$ is achieved: $\pm 0.1$, $\pm 0.2$ and $\pm 0.3$ (dashed curves in red, orange and green). We see that the JC models still outperform PPA till an accuracy on $s$ of around $\pm 0.2$.}
	\label{fig:time}
\end{figure}

While a detailed analysis is beyond the scope of the current work and would require fixing specific experimental parameters, it is worth noticing that, excluding the high temperature regime, the JC protocol appears to be superior to the ideal PPA protocol with 2 ancillas even taking into account two kinds of time limitation:
\begin{enumerate}
	\item Limited waiting time in the cavity, i.e. a bound on the maximum available $s$;
	\item Limited timing accuracy, i.e. the achieved  $s$ fluctuates around a target value.
\end{enumerate}
As a case study, set $\bar{\beta} =1$ and limit $s$ to $s = gt \leq 10$. Then the best available approximation to the $\beta$-swap is realized for $s \approx 7.87$, with the performance monotonically decreasing in a neighbourhood of this value. We then consider an accurate `time-limited JC model' where we set $s= 7.87$; and on top of this we allow various degrees of inaccuracy: errors on $s$ of $\pm 0.1$, $\pm 0.2$ and $\pm 0.3$.  The result are summarized in Fig.~\ref{fig:time}.
The time-limited JC performs better than PPA with 2 ancillas. Adding timing errors on $s$, we obtain a worst-case cooling performance above PPA till around $\pm 0.2$.

\subsection{Jaynes-Cummings model without refreshing the thermal mode}\label{ap:JCreuse}

In the previous result the mode is always reset back to the thermal state after every step. Theorem \ref{th:reusability} shows that when using $U^{\beta}_{SB}$ the same mode can be used repeatedly. Does this still hold (at least approximately) when the interaction is via the Jaynes-Cummings model? 

We explore this question via a numerical simulation of a modified  algorithm where now, at each round $k$, instead of rethermalizing the mode completely, the bosonic mode is partially re-thermalized via a dissipation process.
This is done with a standard master equation, which models the evolution of state $\rho_B$ of the cavity mode due to the interaction with an external thermal field~\cite{scala2007microscopic}:
\begin{align}
\frac{\text{d}\rho_B}{\text{d}t}=-i E \left[a^\dagger a, \rho_B\right] -\frac{1}{2}A n \left[a a^\dagger \rho_B -2 a^\dagger \rho_B a + \rho_B a a^\dagger\right] -\frac{1}{2}A\left(n+1\right)\left[a^\dagger a \rho_B-2a\rho_B a^\dagger +\rho_B a^\dagger a\right].
\label{eq:masterequationcavity}
\end{align}
Here $A$ is the rate of loss of cavity photons (controlling the strength of the re-thermalization) and $ n=\frac{1}{e^{\beta E}-1}$ is the average number of reservoir quanta with energy $E$.

\begin{figure}[t!]
	\centering
	\includegraphics[width=	0.7\linewidth]{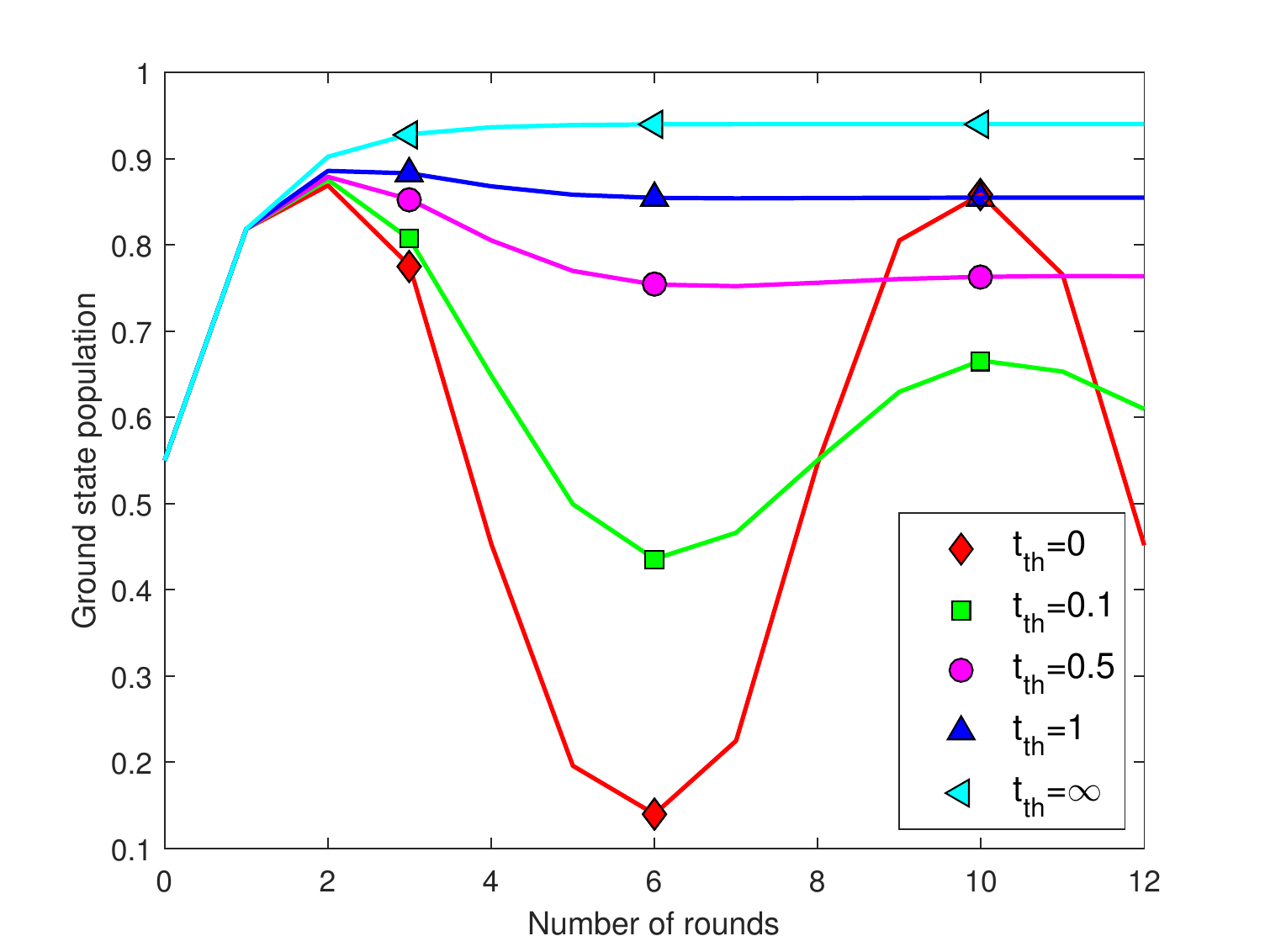}
	\caption{Cooling achieved by the Pauli/Jaynes-Cummings protocols as a function of the number of rounds, when the mode is re-thermalized for various times $t_{\text{th}}$. The infinite time corresponds to the mode being completely re-thermalized at every step (light blue curve at the top), while $t_{\text{th}}=0$ corresponds to no re-thermalization at all (red curve with wide oscillations). We see that re-thermalization is needed to achieve the greatest cooling. For this figure, the parameters are $\beta E=1$, $A=1$, $g=1$ and the Jaynes-Cummings interaction is turned on for a period of time $\tilde{t}=98.92$.}
	\label{fig:plotJC}
\end{figure}

We compare the cooling achieved after $k$ rounds in the case of full reset at each round (Eq.~\eqref{eq:coolingrobustness} with $\epsilon = 1-G_{0|1}(\tilde{s})$), with the cooling achieved for various finite re-thermalization times. In each case, we fix $\beta E =1$ and a particular interaction time between the qubit and the cavity mode, $\tilde{s}=g\tilde{t} = 98.92$. Note that the same procedure can be applied for any $s$, or even taking $s$ to be a random variable to simulate imperfections in the timing. The results are shown in Fig.~\ref{fig:plotJC}. We find that, unlike in the case of implementing $U^{\beta}_{SB}$, one has to reset the mode back to the thermal state as much as possible in order for the algorithm to work efficiently with a fixed interaction time.

Reasonably high cooling can be achieved by interrupting the protocol after $2$ rounds. The above findings suggest the experimental setup discussed in the main text, where atoms are slowly fired inside two identical cavities resonant with the qubit transitions we are cooling. The protocol on each atom then consists of:
\begin{enumerate}
	\item First Pauli $X$ applied.
	\item First JC interaction applied, for some time $\tilde{s}$.
	\item Second Pauli $X$ applied.
	\item Second JC interaction applied, for the same time $\tilde{s}$.
	\item The cavity modes undergo re-thermalization for a finite time according to Eq.~\eqref{eq:masterequationcavity}. 
\end{enumerate}
While in the first step the JC interaction achieves a de-excitation probability of $G_{0|1}(\tilde{s})$ given by Eq.~\eqref{eq:JClambda}, every subsequent atom interacts with only partially re-thermalized cavities, for which  Eq.~\eqref{eq:JClambda} does not hold. Thus, the protocol  may not achieve the same cooling on every atom. Nevertheless, one may expect that, by firing the atoms slowly enough, the re-thermalization of the cavities due to losses will be sufficient to make the cooling performance almost constant. This intuition is confirmed in Fig.~\ref{fig:JC2cavity}. We take the atoms to be initially in a thermal state with $\beta E = 1$. We then plot the final ground state population achieved by each atom passing through the two cavities, as a function of the number of atoms already cooled. The various curves represent different choices for the ratio $A/r$ between the strength of the re-thermalization and the rate at which the atoms are fired (with the caveat that we assume $r$ small enough so that two atoms are never present at the same time in a cavity). When $A/r= \infty$, the single mode has time to re-thermalize perfectly, the performance is the same for each atom and given by Eq.~\eqref{eq:coolingrobustness} with $k=2$ and $\epsilon = 1-G_{0|1}(98.92)$. In realistic scenarios, however, we see that the incomplete thermalization negatively impacts upon the performance. However, we also see that the performance stabilizes to a constant after a small number of atoms are fired, so for a high enough ratio $A/r$ cooling of any number of atoms is possible. This may be understood as the creation of a steady state in the cavity field.
\\
\begin{acknowledgments}
	The authors would like to thank Philippe Faist, Amikam Levy, Mohammad Mehboudi, Nayeli A. Rodr\'iguez-Briones, Raam Uzdin and Marcus Huber for useful discussions and feedback.
	Research  at  Perimeter  Institute  is  supported  by  the
	Government of Canada through the Department of Innovation, Science and Economic Development and by the Province of Ontario through the Ministry of Research, Innovation and Science. ML acknowledges financial support from the the European Union's Marie Sklodowska-Curie individual Fellowships (H2020-MSCA-IF-2017, GA794842), Spanish MINECO (Severo Ochoa SEV-2015-0522 and project QIBEQI FIS2016-80773-P), Fundacio Cellex and Generalitat de Catalunya (CERCA Programme and SGR 875). CP acknowledges financial support from the European Research Council (ERC Grant Agreement no. 337603) and VILLUM FONDEN via the QMATH Centre of Excellence (Grant no. 10059).
\end{acknowledgments}

\bibliographystyle{unsrtnat}	
\bibliography{References}

\end{document}